\documentclass[10pt,twocolumn,aps,pra,superscriptaddress,amsmath,showpacs,tightenlines,pdflatex,longbibliograph]{revtex4-2}
\usepackage[english]{babel}
\usepackage{fontenc}
\usepackage{comment}
\usepackage{amsmath,amsthm,amssymb,amsfonts, bbm}
\usepackage{graphicx,wrapfig,lipsum}
\usepackage{mathtools}
\usepackage{braket}
\usepackage{cases}
\usepackage[dvipsnames]{xcolor} % Gets "DarkGreen" and other colors

\definecolor{forestGreen}{rgb}{0.06, 0.38, 0.06}

\definecolor{burgundy}{rgb}{0.7, 0.0, 0.05}

\usepackage[colorlinks=true,citecolor=burgundy,linkcolor=forestGreen, urlcolor=burgundy]{hyperref}
\usepackage{float}
\usepackage{dsfont}
\usepackage{cleveref}
\newtheorem{thm}{Theorem}[]

\theoremstyle{definition}
\newtheorem{defi}{Definition}
\usepackage{tikz} % TikZ
\usetikzlibrary{calc, arrows, patterns, shapes, decorations, arrows.meta, fit, positioning}
\tikzset{
    auto,node distance =1 cm and 1 cm,semithick,
    var/.style = {minimum width = 0.5cm},
    intvar/.style = {circle, draw, minimum width = 0.5cm, double},
    latent/.style = {minimum width = 0.5cm},
    point/.style = {circle, draw, inner sep=0.06cm, fill, node contents={}},
    triangle/.style = {regular polygon, regular polygon sides=3, draw, inner sep=0.06cm, fill, node contents={}},
    bidir/.style={Latex-Latex,dashed},
    dir/.style={-Latex, thick},
    el/.style = {inner sep=2pt, align=left, sloped}
}
\tikzstyle{vertex}=[circle, fill=black!10, draw=black]
\tikzstyle{edge}=[thick]
\tikzstyle{clique}=[line width=4, draw=black!70]

\usepackage[draft,author=]{fixme}
\fxusetheme{color}

\newcommand{\pedro}[1]{\textcolor{purple}{#1}}

\begin{document}

\title{Latent splitting as a causal probe}
\author{Santiago Zamora}
\affiliation{International Institute of Physics, Federal University of Rio Grande do Norte, 59078-970, Natal, Brazil}
\affiliation{Departamento de F\'isica Te\'orica e Experimental, Federal University of Rio Grande do Norte, 59078-970 Natal, Brazil}
\author{Pedro Lauand}
\affiliation{Perimeter Institute for Theoretical Physics, Waterloo, Ontario, N2L 2Y5, Canada}
\affiliation{Institute for Quantum Computing, University of Waterloo, Waterloo, Ontario, N2L 3G1, Canada}
\author{Isadora Veeren }
\affiliation{ Inria Paris-Saclay, Bâtiment Alan Turing, 1, rue Honor\'e d’Estienne d’Orves—91120 Palaiseau, France}
\affiliation{CPHT, LIX, Ecole polytechnique, Institut Polytechnique de Paris, Route de Saclay—91128 Palaiseau, France}

\author{Davide Poderini}
\affiliation{Università degli Studi di Pavia, Dipartimento di Fisica, QUIT Group, via Bassi 6, 27100 Pavia, Italy}
\author{Rafael Chaves}
\affiliation{International Institute of Physics, Federal University of Rio Grande do Norte, 59078-970, Natal, Brazil}
\affiliation{School of Science and Technology, Federal University of Rio Grande do Norte, Natal, Brazil}

\date{\today}	
\begin{abstract}
%Understanding the causal origins of observed correlations is a central goal of empirical science. Bell’s theorem shows that quantum theory is incompatible with classical causal models, a phenomenon known as quantum nonlocality. 

Generalizations of Bell’s framework to causal networks have yielded new foundational insights and applications, including the use of interventions to enhance the detection of nonclassicality in scenarios with communication. Such interventions, however, become uninformative when all observable variables are space-like separated. To address this limitation, we introduce the \emph{latent splitting} procedure, a generalization of interventions to quantum networks in which controlled manipulations are applied to latent quantum systems. We show that latent splitting enables the detection of nonclassicality by combining observational and interventional data even when conventional interventions fail. Focusing on the triangle network, we derive new analytical witnesses that robustly certify nonclassicality, including nonlinear inequalities for minimal binary-variable scenarios and extensions of the nonclassical region of previously proposed experiments.
\end{abstract}
\maketitle

\section{Introduction}
  
%\fixthis{Check, edit and add more relevant references}

The central aim of science is to uncover relationships between natural phenomena by formulating hypotheses that explain the often hidden causal mechanisms underlying observable events. In this context, Bell’s theorem \cite{bell1964einstein} revealed a fundamental incompatibility between the predictions of quantum mechanics and any classical theory based on local hidden variables. Owing to its minimal set of assumptions, the Bell framework provides particularly strong guarantees for information-processing tasks, including device-independent randomness certification \cite{acin2012randomness}, quantum key distribution (QKD) \cite{ekert1991quantum,acin2007device,nadlinger2022experimental} and other cryptographic protocols \cite{sen2003unified,moreno2020device}, as well as communication-complexity problems \cite{brukner2004bell,buhrman2010nonlocality}. Beyond these applications, the violation of Bell inequalities remains a cornerstone of the foundations of quantum theory \cite{brunner2014bell}.

It is therefore natural that considerable effort has been devoted to extending and generalizing the standard Bell scenario \cite{tavakoli2022bell}. From a practical perspective, rapid experimental progress motivates the development of theoretical frameworks capable of supporting increasingly complex quantum technologies \cite{simon2017towards,wehner2018quantum,dynes2019cambridge,chen2021integrated}. These include large-scale quantum communication networks \cite{carvacho2017experimental,saunders2017experimental,poderini2020experimental,gu2023experimental,wang2024experimental}, architectures for distributed quantum computing \cite{wei2019experimental,ho2022entanglement}, quantum clock synchronization protocols \cite{komar2014quantum}, and related applications. From a foundational viewpoint, exploring network scenarios beyond Bell has revealed novel features: the indispensability of complex numbers in quantum theory \cite{renou2021quantum}, forms of non-classicality that do not rely on measurement choices \cite{fritz2012beyond,renou2019genuine, Abergsemi2020,chaves2021causal,polino2023experimental,lauand2023witnessing}, refined notions of multipartite nonlocality \cite{pozas2022full,suprano2022experimental}, and new avenues for the self-testing of physical theories \cite{weilenmann2020self}.

Despite this progress, most existing results focus exclusively on observational data generated by a quantum network, namely on correlations between measurement outcomes at different nodes. While such correlations can demonstrate the incompatibility of quantum predictions with classical causal explanations, they neglect a central concept from causal inference: interventions \cite{pearl2009causality,korb2004}. An intervention consists of actively setting a variable to a chosen value, thereby severing its causal dependence on other variables in the network. The additional information obtained through such controlled manipulations can substantially strengthen causal tests and reveal non-classical features in scenarios that would otherwise admit a classical explanation based solely on observational data \cite{gachechiladze2020quantifying,agresti2022experimental,lauand2024quantum,lauand2024quantum2}.

While interventions on observed variables play a central role in classical causal modeling, a fundamental distinction between classical and quantum descriptions of latent variables leads to markedly different possibilities in quantum networks. In classical causal models, latent variables represent uncontrolled or unobserved external factors and are, by definition, inaccessible to experimental manipulation. In quantum causal models, by contrast, latent nodes correspond to intrinsically nonclassical degrees of freedom, typically quantum systems, that mediate correlations between observed variables. Crucially, unlike their classical counterparts, such quantum latent systems can be prepared, transformed, and controlled using well-established experimental techniques \cite{Nielsen_Chuang_2010,serafini2017quantum}.

Motivated by this observation, we introduce the notion of \emph{latent splitting}, defined as a type of intervention applied not to observable nodes but to an edge of a latent node in a quantum causal network. We develop a general novel framework for such interventions and demonstrate that they provide a powerful new tool for causal modeling, capable of revealing causal and operational features of a scenario that remain inaccessible to standard Pearl interventions. To illustrate the strength of this approach, we analyze several aspects of the triangle scenario, the simplest nontrivial causal structure where latent splitting becomes meaningful and where traditional interventions are inherently uninformative.

The paper is organized as follows. In Sec.~\ref{sec: preliminaries}, we review the essential background on causal modeling with directed acyclic graphs, the triangle network scenario and the standard Pearl-like interventions. Section~\ref{sec: Partial_interventions} presents our main contribution: the framework of latent splitting. In Section~\ref {sec: results}, we first demonstrate how the proposed framework encompasses standard Pearl-like interventions. We then apply it to the triangle network and present two main results. First, by combining  latent splitting with the inflation technique \cite{WolfeSpekkensFritz_2019}, we extend the range of nonclassicality of the RGB4 family of probability distributions \cite{renou2019genuine}, demonstrate robustness to small amounts of noise, and derive polynomial inequalities capable of witnessing this behavior. Second, we show that latent splitting enables the certification of nonclassicality in the minimal triangle scenario, in which all observable variables are binary, by explicitly constructing a nonlinear Bell inequality. This resolves an open question in the literature \cite{boreiri2023towards,da2025local} and, furthermore, we show this inequality can only be violated if we consider interventional data. Finally, Sec.~\ref{sec: conclusion} concludes and outlines future research directions.

\section{Preliminaries}\label{sec: preliminaries}
This section introduces the basic concepts of causal modeling \cite{pearl2009causality}, the triangle scenario \cite{fritz2012beyond}, and interventions. We also establish the notation used throughout the manuscript.

\subsection{Causal modeling}
The derivation of a Bell inequality can be viewed as a specific instance of a more general problem in causal inference: determining whether an observed behavior is compatible with a given causal structure. This task is known as \emph{causal compatibility}. In the context of quantum networks, causal modeling~\cite{pearl2009causality} provides a natural framework to represent such scenarios as \emph{directed acyclic graphs} (DAGs). In this representation, measurement settings and outcomes are treated as observable classical variables, while the sources are modeled as latent variables. The central question is whether the correlations generated within a given network configuration --namely, the probability distributions of outputs conditioned on inputs-- can be reproduced by the corresponding classical causal model.

Formally, a DAG is a graph $G=(V,E)$ defined by a finite set of nodes (or vertices) $V$ and a set of directed edges $E \subseteq V \times V$, where each edge is an ordered pair $(v_1, v_2)$. In causal modeling, each node $v \in V$ represents a random variable. We distinguish between observable variables ($V_{O}$), denoted by Latin letters (e.g., $A, B$), and latent variables ($V_{L}$), denoted by Greek letters (e.g., $\alpha, \beta$). Directed edges encode direct causal relations: an edge $A \to B$ indicates that $A$ is a direct cause of $B$. The acyclic property ensures that no variable is a cause of itself. An illustrative example of a DAG is shown in Fig.~\ref{fig:examples-DAG}. While disconnected nodes do not directly influence one another, they may nonetheless be correlated through common causes.

Consider a graph $G$ with observable variables $A_1,\ldots,A_n \in V_{O}$ and latent variables $\Lambda_1,\ldots,\Lambda_m \in V_{L}$. The graph $G$ encodes the causal structure of the network through the notion of parenthood. For any node $X \in V$, we define its set of direct causes, or \emph{causal parents}, as $\text{Pa}(X) := \{Y \in V \mid Y \to X \}$. Conversely, the set of direct effects, or \emph{causal children}, is defined as $Ch(X) := \{Y \in V \mid X \to Y \}$. 

A classical causal model associated with $G$ consists of a collection of conditional probability distributions (causal parameters) $p(X \mid \text{Pa}(X))$ for each node $X \in V$. Whenever $\text{Pa}(X)=\emptyset$, the corresponding distribution reduces to $p(X)$. Throughout this work, we assume that $\text{Pa}(\Lambda_j)=\emptyset$, i.e., latent variables have no causal parents. A probability distribution $P(a_1,\dots,a_n)$ over the observable variables is said to be classical with respect to $G$ if there exists a set of causal parameters such that it admits the decomposition
\begin{equation}
    P(a_1,\dots,a_n) = \sum_{\lambda_1..\lambda_m} \prod_{i=1}^m p(\lambda_i)\prod_{j =1}^n p(a_j|\text{Pa}(a_j)).
    \label{eq:MarkovCondition}
\end{equation}

If the sources instead distribute quantum systems, the latent variables $\Lambda_1,\dots,\Lambda_m$ are represented by normalized \emph{density matrices} $\rho_{\Lambda_1},\dots,\rho_{\Lambda_m}$. To each observable variable $A_1,\dots,A_n$ we associate a positive operator-valued measure (POVM), described by operators $E_{a_j|\text{Pa}^O(a_j)}$, where $\text{Pa}^O(A_j):=\{Y\in V_O \mid Y \rightarrow A_j\}$ denotes the observable parents of $A_j$. The Hilbert space of each source $\rho_{\Lambda_i}$ is labeled by the set of its children variables, $\text{Ch}(\rho_{\Lambda_i}):=\{Y\in V \mid \Lambda_i \rightarrow Y\}$, and each operator $E_{a_j|\text{Pa}^O(a_j)}$ acts nontrivially on the Hilbert space associated with party $A_j$. 

A probability distribution $P_Q(a_1,\dots,a_n)$ is said to admit a \emph{quantum causal model} if it can be written as
\begin{equation}
        P_Q(a_1,\dots,a_n) = \text{Tr}\!\left( \bigotimes_{i=1}^m \rho_{\Lambda_i} \bigotimes_{j=1}^n E_{a_j|\text{Pa}^O(a_j)} \right),
    \label{eq:Quantum_MarkovCondition}
\end{equation}
where the operators satisfy $\sum_{a_j}E_{a_j|\text{Pa}^O(a_j)}=\mathbf{1}$ and $\text{Tr}(\rho_{\Lambda_i})=1$.

\subsection{The triangle scenario}
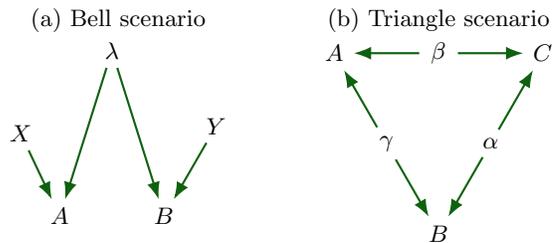
\begin{figure}[t]

\begin{minipage}[t]{0.48\columnwidth}
    \centering
    (a) Bell scenario
    \begin{tikzpicture}
        \draw (120 - 30: 1.8cm) node[latent] (l1) {$\lambda$};
        \draw (120 + 30: 1.4cm) node[var] (x) {$X$};
        \draw (240 - 30: .8cm) node[var] (a) {$A$};
        \draw (360 - 30: .8cm) node[var] (b) {$B$};
        \draw (360 + 30: 1.6cm) node[var] (y) {$Y$};

        \path[dir, forestGreen] (l1) edge (a) (l1) edge (b) (x) edge (a) (y) edge (b);
    \end{tikzpicture}

\end{minipage}
\begin{minipage}[t]{0.48\columnwidth}
    \centering
    (b) Triangle scenario
    \begin{tikzpicture}

        \draw (120 - 30: .8cm) node[latent] (l1) {$\beta$};
        \draw (120 + 30: 1.6cm) node[var] (a) {$A$};
        \draw (240 - 30: .8cm) node[latent] (l2) {$\gamma$};
        \draw (240 + 30: 1.6cm) node[var] (c) {$B$};
        \draw (360 - 30: .8cm) node[latent] (l3) {$\alpha$};
        \draw (360 + 30: 1.6cm) node[var] (b) {$C$};

        \path[dir, forestGreen] (l1) edge (a) (l1) edge (b); 
        \path[dir, forestGreen] (l2) edge (c) (l2) edge (a);
        \path[dir, forestGreen] (l3) edge (b) (l3) edge (c);

    \end{tikzpicture}

\end{minipage}
\caption{\textbf{DAGs describing Bell and triangle scenarios.} Roman letters denote observed variables, while greek letters denote latent variables. (a)~In a bipartite Bell scenario, a latent variable $\lambda$ influences the outcomes of two parties, with inputs $X,Y$ and outputs $A,B$. Notice that the inputs are just a particular instance of observed variables. (b)~In the triangle scenario, three independent sources, given by latent variables $\alpha, \beta, \gamma$, influence the observed variables $A,B,C$. }
\label{fig:examples-DAG}
\end{figure}

A key insight emerging from generalizations of Bell’s theorem to causal networks is that quantum nonclassicality can arise solely from the network structure of independent sources. In particular, quantum nonclassicality can be demonstrated even in the absence of measurement inputs \cite{Branciard2012,fritz2012beyond,boreiri2023towards}, meaning that all observers perform a single, fixed measurement—an outcome in stark contrast with the standard Bell scenario. Prominent examples of such input-free nonclassicality have been identified in the triangle network \cite{renou2019genuine,Renou2022,Abiuso2022,boreiri2023towards}, depicted in Fig.~\ref{fig:examples-DAG}(b).

The triangle network has attracted considerable recent interest, serving as a testbed for novel forms of entanglement \cite{navascues2020genuine} and nonlocality \cite{renou2019genuine}, as well as for demonstrating the power of the inflation technique \cite{pozas2023proofs,gitton2025}. It consists of three parties --Alice, Bob, and Charlie-- and three independent bipartite sources $\alpha$, $\beta$, and $\gamma$, arranged such that each pair of parties shares exactly one source. 

The set of probability distributions that are classically compatible with the triangle network is given by
\begin{align}\label{eq:triangle-markov}
P(a,b,c)=\sum_{\alpha,\beta,\gamma}p(\alpha)p(\beta)p(\gamma)p(a|\beta,\gamma)p(b|\alpha,\gamma)p(c|\alpha,\beta),
\end{align}
where $p(\alpha)$, $p(\beta)$, and $p(\gamma)$ denote the probability distributions associated with the three sources, and $p(a|\beta,\gamma)$, $p(b|\alpha,\gamma)$, and $p(c|\alpha,\beta)$ are the local response functions of Alice, Bob, and Charlie, respectively.

In the quantum case, the set of compatible probability distributions in the triangle scenario is given by the Born rule,
\begin{equation}
    P_Q(a,b,c)=\text{Tr}\!\left(E_a\otimes E_b\otimes E_c \left(\rho_{\gamma}\otimes\rho_{\beta}\otimes\rho_{\alpha}\right) \right),
\end{equation}
where $\rho_{\gamma}$, $\rho_{\beta}$, and $\rho_{\alpha}$ represent the quantum states distributed by the sources $\gamma$, $\beta$, and $\alpha$, respectively, and $E_a$, $E_b$, and $E_c$ are POVM elements satisfying $E_i\succeq 0$ and $\sum_i E_i=\mathds{1}$ for $i\in\{a,b,c\}$. The characterization of quantum correlations in this scenario is particularly challenging, largely due to the assumption of source independence.

%In the next section, we briefly review the inflation technique, a general and powerful method for addressing this challenge and detecting nonclassicality in quantum networks.

\subsection{Interventions}
\label{subsec: pearl_interventions}

Interventions are a central tool in the causal inference framework, serving as a key mechanism for understanding the causal relationship between variables \cite{pearl2009causality}. In the context of causal inference, an intervention is any action taken by the experimenter to change a variable systematically, such that it is possible to observe the effects associated with some of its outcomes, eliminating any bias from its natural confounding causes. In the following, we will see different approaches for studying interventions and we start by considering \emph{Pearl-like} interventions.

Statistics under a Pearl-like intervention are formally described by  \emph{do-probabilities}. A do-probability quantifies the likelihood of an event after an external intervention modifies the system. This intervention is modeled by excising the variable's natural causal mechanism from the model and replacing it with a specific, fixed value. These operations, often termed \emph{Pearl-like} interventions, apply exclusively to the observed variables within the network. As such, they can be considered node interventions.
 
One of the primary examples that showcase the importance of Pearl-like interventions in modern science are medical trials \cite{Bikak_2019,Gani2023}. Medical trials are designed to certify causality between treatment, usually implemented by a drug, and the patient's recovery. In this setup, we can consider binary random variables $A,B$, where $A$ corresponds to administering a drug or not, if $a=0$ or $a=1$, respectively, and $B$ similarly corresponds to the outcome of the patient, for example, whether the patient recovers from the illness. In this context, the probability $P(a,b)$ represents the joint probability of the treatment being administered and the recovery occurring, and it is well known that a correlation between treatment and recovery is not enough to infer a cause-and-effect relationship between them.
If we observe a correlation, that is, $P(a,b)\neq P(a)P(b)$, the most general causal model explaining such correlations might involve direct influences and confounding factors, denoted by $\lambda$, which for a variety of reasons might not be directly observed. These reasons can include age, genetics, preexisting medical conditions, and others. Formally, this means that the conditional observational distribution $P(a,b)$ can be decomposed as 
\begin{equation}  P(a,b)=\sum_{\lambda}p(\lambda)p(a|\lambda)p(b|\lambda,a).
\end{equation}
To ensure that any differences in recovery between the two groups can be attributed to treatment rather than confounding factors, participants are randomly assigned to the treatment group (in which they are forced to receive treatment) or to the control group (in which they receive a placebo). This experiment implements an intervention where we can represent the probability of recovery ($B$) given that the treatment ($A$) was administered as $P(b|do(a))$, where the ``$do(a)$" notation indicates that the treatment was actively applied, as opposed to simply observing the treatment happen naturally. The \emph{do-conditional} $P(b|do(a))$ isolates the causal impact of the treatment on the recovery by eliminating confounding factors, \textit{i.e.} it can be written as 
\begin{equation}
    P(b|do(a))=\sum_{\lambda}p(\lambda)p(b|\lambda,a),
\end{equation}
essentially erasing the influence of $p(a|\lambda)$ from the model. In the classical regime, confounding factors play the role of the variables that are inaccessible to the experimenter, and, thus, to leverage interventional data there must be a direct causal influence between the observed variables in the experiment.

Moreover, in the context of quantum networks, \textit{i.e.} when latent variables are described by a density operator, one can analogously define Pearl-like interventions. We formalize this notion  in Appendix \ref{app: thm_1}.
%\begin{defi}[Pearl intervention] \label{def: PearlInt} Given a network of $n$ parties,  $A_1,\dots,A_n$, with $m$ independent sources, $\Lambda_1,\dots,\Lambda_m$, where each source distributes states $\rho_i$. A Pearl intervention is defined relative to one of the $n$ parties of the network. Therefore, given a party $A_l\in V_O$, a Pearl intervention performed on $A_l$ is represented by the do-conditional $p(a_1,\dots,a_{l-1},a_{l+1},\dots,a_n|do(a_l))$,  given by 
%\begin{equation}\label{eq: pearl_do}
%P(a_1,\dots|do(a_l))=\text{Tr}\left(\bigotimes_{i}\rho_i\left[\left(\bigotimes_{ j\neq l}E_{a_j|Pa^O(a_j)}\right)\otimes \mathbf{1} \right]\right)
%\end{equation}
%where $p(a_1,\dots|do(a_l))$ takes statistics over all nodes $A_j\in V_O/\{A_l\}$. The definition essentially replaces the operator $E_{a_l|\text{Pa}^O(a_l)}$ by an identity $\mathbf{1}$, and leaves the remaining objects unchanged. 
%Equivalently, Eq.~(\ref{eq: pearl_do}) can be rewritten using the partial trace operation as 
%\begin{equation}
%    p(a_1,\dots|do(a_l))=\text{Tr}\left(\left[\text{Tr}_{A_l}\left(\bigotimes_i \rho_i\right)\right]\bigotimes_{j\neq l} E_{a_j|\text{Pa}^O(a_j)}\right).
%\end{equation}
%\end{defi}
Notably, Pearl-like interventions have recently been shown to be a powerful tool in detecting nonclassicality in scenarios with communication where the measurement outcomes of one party may influence the choice of measurement basis of another \cite{chaves2018quantum,gachechiladze2020quantifying,hutter2023quantifying}. In such cases, we may define an alternative experiment in which we force the choice of the measurement basis of the party that was originally influenced by the outcome of the other and, from this setup, estimate the \emph{do-conditional} probability. This extra data set allows us to certify quantum nonclassicality in a new data regime, which may show significant advantages such as better threshold detection efficiencies~\cite{poderini2024observational} and noise robustness~\cite{lauand2024quantum}.

However, in scenarios like the triangle network, where there are no direct causal influences between the observed nodes, Pearl-like interventions would not give us any new information other than the observational distribution. To see this, one can consider the graph shown in Fig.~\ref{fig:examples-DAG}b) and fix the variable $A$, the resulting $do$-probability is
\begin{equation}
    P(b,c|do(a)) := \sum_\alpha p(\alpha)p(b|\alpha)p(c|\alpha)=P(b,c),
\end{equation}
where $P(b,c)$ is just the marginal of the observational distribution written in Eq.~(\ref{eq:triangle-markov}). Then a natural question becomes: How to include interventions in scenarios where all variables are space-like separated? 

In the next section, we propose a generalization of Pearl-like interventions for quantum networks answering this question. We consider some examples in the triangle network and show that this generalization allows for a stronger certification of nonclassicality.

\section{Latent Splitting}\label{sec: Partial_interventions}

In this sectio we propose a new framework for interventions called a \emph{latent splitting}, which can be understood as a natural generalization of a Pearl-like intervention. Notably, we will show in the following sections that we can always estimate all do-conditionals available in a network by performing latent splitting; however, the converse is not true: there are cases where latent splitting  reveal information that Pearl interventions cannot. %We show this by considering some examples in the triangle network.

A key distinction between classical and quantum causal models lies in the nature of their latent variables. In the classical regime, latent variables are typically defined as inaccessible properties used to explain observed correlations. Causal inference is then concerned with interventions on the observables in the presence of these unobserved confounders. In quantum networks, however, latent variables represent the quantum state $\rho$, a physical property. While not directly measured, this state is experimentally controllable. The key difference, therefore, is the possibility of performing interventions directly on the latent source itself (e.g., by manipulating the preparation of $\rho$), which expands the class of possible interventions beyond those considered on the observables alone.

%In the classical regime, a latent variable represents properties whose existence is inferred from the correlations between observable variables and are inaccessible to the experimenter. Consequently, if one wants to leverage interventional data, there must be a direct causal influence between the observed variables in the experiment. However, in the context of quantum information, latent variables are used to model \emph{physical properties}, which can only be measured indirectly and are described by some state $\rho$. Although such physical properties may not be measured directly, they can still be manipulated in the laboratory. Think for example in a continuous variable system, in quantum optics \cite{serafini2017quantum}. The key observation here hinges precisely on this crucial fact that latent variables are not completely inaccessible in the context of quantum networks and, therefore, make it possible for an intervention to be performed.

A simple example is to consider a source producing two photons (for example, through spontaneous parametric down-conversion (SPDC)): one distributed to Alice and the other to Bob. In this setup, the state of the photons may be, in principle, unknown to both Alice and Bob. Although the state may be unknown, Alice has a certain control on the states sent by the source: for example, she can still block the path of her photon and replace it by preparing any other state of her knowledge. This is an example of what we call \emph{latent splitting}: it is the action in which a party discards its shares of certain incoming quantum states and replaces them with new locally prepared states of its own. In the following sections, we show that latent splitting can be understood as a generalization of Pearl-like interventions.

Here, we will consider two-layer networks composed of $n $ parties,  $A_1, \dots,A_n$, and $m$ independent sources, $\Lambda_1,\dots,\Lambda_m$, where each source distributes states $\rho_{\Lambda_i}$, or simply $\rho_i$. And, for our purposes, we will consider the case where Alice simply re-prepares the source of the experiment locally. 

\begin{defi}[Latent splitting]\label{def: ParInt}
The \emph{latent splitting} procedure is defined relative to an edge of one of the $m$ sources of the network. Therefore, given an edge of the network, which connects the source $\Lambda_k\in V_L$ to a party $A_l\in V_O$, latent splitting is the action of discarding the share of $\Lambda_k$ that goes to $A_l$ and re-preparing an independent share $\hat{\rho}_{k,l}:=\text{Tr}_{V_O/\{A_l\}}(\rho_{obs})$. As a result, the global state of the original network $\rho_{obs}:=\bigotimes_{i}\rho_{i}$ is changed to a new state $\rho_{int}^{(k,l)}$ given by 
\begin{equation}
     \rho_{int}^{(k,l)}:=\left(\hat{\rho}_{k,l}\otimes \text{Tr}_{A_l}(\rho_k)\right)\otimes\left(\bigotimes_{i\neq k }\rho_{i}\right).
\end{equation}
By collecting the corresponding statistics of this new experiment we will have
\begin{equation}
\label{eq:interventionalP}
    P^{(k,l)}(a_1,\dots,a_n):=\text{Tr}\left(\rho^{(k,l)}_{int}\bigotimes_j E_{a_j|Pa_O(a_j)}\right).
\end{equation}
This formalizes the procedure of latent splitting performed on an edge of a latent variable in the quantum network. This concept naturally extends to scenarios where multiple parties may simultaneously perform multiple latent splittings on their respective
subsystems by consecutively applying this definition for each latent splitting procedure.
\end{defi}

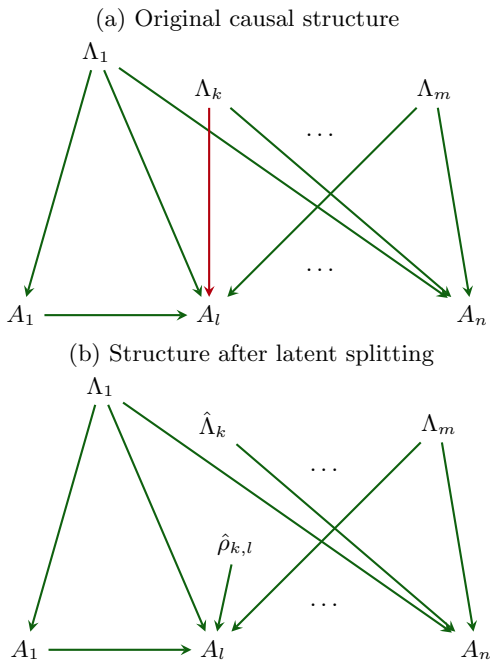
\begin{figure}[t]
    \centering
    (a) Original causal structure
    \tikzset{
        var/.style={}, 
        latent/.style={},
        dir/.style={->, >=stealth, thick}
    }

    % --- Subfigure (a): The Original Network ---
    % Allow the subfigure to take up most of the column width
    \begin{minipage}[t]{1\columnwidth}
        \centering
        \begin{tikzpicture}
            % Define m sources schematically
            \node[latent] (l_j) at (0, 3) {$\Lambda_k$};
            \node[latent] (l_1) at (-1.5, 3.5) {$\Lambda_1$};
            \node at (1.5, 2.4) {$\dots$};
            \node[latent] (l_m) at (3, 3) {$\Lambda_m$};
            
            % Define n parties schematically
            \node[var] (x_k) at (0, 0) {$A_l$};
            \node[var] (x_1) at (-2.5, 0) {$A_1$};
            \node at (1.5, 0.6) {$\dots$};
            \node[var] (x_n) at (3.5, 0) {$A_n$};

            % General connections
            \path[dir, forestGreen, thick] (l_1) edge (x_1);
            \path[dir, forestGreen, thick] (l_1) edge (x_n);
            \path[dir, forestGreen, thick] (l_m) edge (x_n);
            \path[dir, forestGreen, thick] (l_j) edge (x_n);
            \path[dir, forestGreen, thick] (x_1) edge (x_k);
            
            % Add multiple connections to X_k
            \path[dir, forestGreen, thick] (l_1) edge (x_k);
            \path[dir, forestGreen, thick] (l_m) edge (x_k);
            
            % Highlight the specific link to be intervened upon
            \path[dir, burgundy,  thick] (l_j) edge (x_k);
        \end{tikzpicture}
        \label{fig:general-network-before}
    \end{minipage}
    
    % A blank line here creates the vertical stacking
    
    % --- Subfigure (b): The Network After Intervention ---
    \begin{minipage}[t]{1\columnwidth}
        \centering
        (b) Structure after latent splitting
        \begin{tikzpicture}
            % Define m sources schematically
            \node[latent] (l_j) at (0, 3) {$\hat{\Lambda}_k$};
            \node[latent] (l_1) at (-1.5, 3.5) {$\Lambda_1$};
            \node at (1.5, 2.4) {$\dots$};
            \node[latent] (l_m) at (3, 3) {$\Lambda_m$};
            \node[latent] (L_hat) at (0.3, 1.4) {$\hat{\rho}_{k,l}$};
            
            % Define n parties schematically
            \node[var] (x_k) at (0, 0) {$A_l$};
            \node[var] (x_1) at (-2.5, 0) {$A_1$};
            \node at (1.5, 0.6) {$\dots$};
            \node[var] (x_n) at (3.5, 0) {$A_n$};

            % General connections
            \path[dir, forestGreen,thick] (l_1) edge (x_1);
            \path[dir, forestGreen,thick] (l_m) edge (x_n);
            \path[dir, forestGreen,thick] (l_1) edge (x_n);
            \path[dir, forestGreen,thick] (l_j) edge (x_n);
            \path[dir, forestGreen,thick] (x_1) edge (x_k);

            % Connections to X_k, but with the red edge removed
            \path[dir, forestGreen,thick] (l_1) edge (x_k);
            \path[dir, forestGreen,thick] (L_hat) edge (x_k);
            \path[dir, forestGreen,thick] (l_m) edge (x_k);
        \end{tikzpicture}
        
    \end{minipage}
    %\begin{subfigure}[b]{1\columnwidth}
        %\centering
        %\begin{tikzpicture}
            % Define m sources schematically
            %\node[latent] (l_j) at (0, 3) {$\lambda_j$};
            %\node[latent] (l_1) at (-1.5, 3.5) {$\lambda_1$};
            %\node at (1.5, 3.5) {$\dots$};
            %\node[latent] (l_m) at (3, 3) {$\lambda_m$};
            %\node[latent] (L_hat) at (-1, 0.6) {$\hat\Lambda_k$};
            
            % Define n parties schematically
            %\node[var] (x_k_numb) at (-1.3, 1.5) {$X^\#_k$};
            %\node[var] (x_k) at (0, 0) {$X_k$};
            %\node[var] (x_1) at (-2.5, 0) {$X_1$};
            %\node at (2, 0) {$\dots$};
            %\node[var] (x_n) at (3.5, 0) {$X_n$};

            % General connections
            %\path[dir, gray] (l_1) edge (x_1);
            %\path[dir, gray] (l_1) edge (x_k_numb);
            %\path[dir, gray] (x_1) edge (x_k_numb);
            %\path[dir, gray] (L_hat) edge (x_k_numb);
            %\path[dir, gray] (x_1) edge (x_k);
            %\path[dir, gray] (l_m) edge (x_n);
            %\path[dir, gray] (l_j) edge (x_n);
            %\path[dir, gray] (l_j) edge (x_k);
            %\path[dir, gray] (l_1) edge (x_n);

            % Connections to X_k, but with the red edge removed
            %\path[dir, gray] (l_1) edge (x_k);
            %\path[dir, gray] (l_m) edge (x_k);
        %\end{tikzpicture}
        %\caption{A joint causal structure}
        
    %\end{subfigure}
    
    \caption{\textbf{Illustration of a latent splitting.} On top (a) is the original causal structure, where party $A_l$ is influenced by several sources. The link from source $\Lambda_k$  to $A_l$ are highlighted in red. In the next panel (b) it is shown the effective causal structure after the intervention, where party $A_l$ has replaced its incoming state from source $\Lambda_k$, effectively severing that causal link.}
    \label{fig: general_patial_interv}
\end{figure}

Notably, we can understand latent splitting defined above graphically. When performing latent splitting, for each intervention, we select an edge of the source $\Lambda_k$ in the original DAG. Latent splitting can be understood as the action that erases this edge from the original causal structure and substitutes it by another latent node, denoted $\hat{\rho}_{k,l}$ with the same outcoming structure, see Fig.~\ref{fig: general_patial_interv}. This is represented in the definition by the independent preparation $\hat{\rho}_{k,l}\otimes\text{Tr}_{A_l}(\rho_k)$ and can be composed sequentially, i.e., we can consider $P^{(k,l)(r,s)...}(a_1,...,a_n)$.

A natural consequence of our definition is that given a causal structure one can define alternative causal structures after splitting that mirrors certain aspects of the original network. In the context of causal inference and network nonlocality, these alternative experiments can be understood as a particular instance of the inflation technique, a subclass of scenarios denominated \emph{non-fanout inflations}. Non-fanout inflations are a class of inflations that are accessible for any general probabilistic theory (GPT), which consists of causal structures composed of independent copies of the variables in the original DAG. The latent splitting procedure establishes that some non-fanout inflations will be experimentally accessible by performing this type of intervention. Namely, the non-fanout inflations that are available via latent splitting are the ones that may have many copies of the sources rewired in the network but only one copy of each observable variable of the original causal structure.

As will be shown in the following, the standard notion of a Pearl-like (node) intervention can be recovered by suitably combining different latent splittings. In this sense, latent splitting generalizes node interventions. Another important distinction arises when comparing latent splittings with measurement inputs. In the case of measurement inputs, the intervention corresponds to modifying the POVM operators $E_{a_j|Pa_O(a_j)}$ as a function of the inputs $Pa_O(a_j)$ in Eq.~\eqref{eq:interventionalP}. That is, the input state remains fixed, while the measurement device is changed. By contrast, latent splitting acts in the opposite way: the measurement device, and hence the POVM operators describing it, are kept fixed, whereas a portion of the input state is modified—specifically, the component associated with the edge on which the latent splitting is performed.

\section{Results}\label{sec: results}
In this section, we present the core results of our framework. We begin by establishing a general equivalence in the context of causal inference: Theorem~\ref{thm:Pearl_from_partial} demonstrates that any Pearl-like intervention can be perfectly recovered by a suitable composition of latent splittings. We then show that while standard interventions are uninformative in space-like separated scenarios like the triangle network, latent splitting provides the necessary leverage to derive analytical witnesses for non-classicality. To demonstrate this power, we integrate our framework with the inflation technique \cite{WolfeSpekkensFritz_2019} to certify non-classicality in two paradigmatic cases: the RGB4 distribution---where we provide robust noise-tolerance bounds---and a binary coarse-grained version of the Fritz-like protocol, which remains classically reproducible without the use of latent splittings.

\subsection{Pearl interventions from latent splitting}

Our first result shows that latent splitting can always estimate Pearl interventions. Operationally, a Pearl intervention on $A_l$ is implemented by removing the mechanism that generates the outcomes of $A_l$, and replacing it by a mechanism that forces the variable $A_l$ to take some value $a_l$, while keeping all the other causal influences intact.
In the following, we show that latent splitting can reproduce this effect and, to show this, we prove that one can always estimate do-conditionals from data originated from latent splittings. We state this result in the following theorem.

\begin{thm}[]\label{thm:Pearl_from_partial}
   Let us consider a two-layered network,  with sources $ \Lambda_1,\dots,\Lambda_m$, each distributing a system $\rho_i$, and parties $A_1,\dots,A_n$, performing measurements $E_{a_j|Pa^O(a_j)}$. Then any do-conditionals on this network, representing a Pearl-like intervention, can always be inferred by data from a collection of latent splittings and the observational distribution.
\end{thm}
We leave the proof of Theorem \ref{thm:Pearl_from_partial} to Appendix \ref{app: thm_1}, and here we give the main ideas of the proof.

\emph{Sketch of proof.} First, we show that by composing all latent splittings that an arbitrary party, $A_l$, can perform on the sources to which it is connected, produces a state $\rho_{int}$ given by 
\begin{equation}
     \rho_{int}= \text{Tr}_{A_l}\left(\rho_{obs}\right)  \otimes \rho^{A_l}_{obs}.
\end{equation}
where $\rho^{A_l}_{obs}$ is simply the portion of $\rho_{obs}$ that is distributed to $A_l$. Then, the statistics of this intervention yields 
\begin{equation}
     P_{int}(a_1,\dots,a_n)=P(a_1,\dots|do(a_l))P(a_l|do(\text{Pa}^O(a_l)).
\end{equation}
If $\text{Pa}^O(A_l)=\emptyset$, then $P(a_l|do(\text{Pa}^O(a_l))\equiv P(a_l)$, and we are done. However, if  $\text{Pa}^O(A_l)\neq \emptyset$, then we must apply this procedure inductively. That is, since the causal structure is acyclic, there must be at least one variable, $A_{l^*}$ such that $\text{Pa}^O(A_{l^*})=\emptyset$, then we can estimate its do-conditional $P(a_1,\dots|do(a_{l^*}))$, and use this do-conditional to estimate the other interventions for variables which are children of $A_{l^*}$. By doing this procedure, we cover all observable variables in the network with finitely many steps and, thus, estimate all Pearl-like interventions.
\qed

Theorem \ref{thm:Pearl_from_partial} shows that all possible do-conditionals that one could consider given a generic network can be expressed as a classical post-processing of $p_{obs}$ and a collection of $p_{int}$ from different latent splittings. Therefore, we can consider latent splitting as a procedure that subsumes the typical notion of interventions over observed variables. In the next sections, we will show how one can leverage latent splitting in the triangle network, where Pearl-like interventions are trivial, and in Appendix \ref{app: partial_equal_Pearl}, we show particular cases where both kinds of interventions coincide. 

\subsection{Quantum nonclassicality with latent splitting}
Early demonstrations of nonlocality in networks often relied on standard bipartite Bell tests, showing that bipartite nonlocality can be effectively “disguised” within a network, thereby giving rise to network-nonclassical distributions~\cite{fritz2012beyond,chaves2021causal}. This line of work naturally raised the question of whether entirely new forms of quantum nonlocality, intrinsically linked to the structure of the network itself, could arise. Subsequent studies have provided evidence that such genuinely network-specific nonlocal correlations can indeed occur \cite{renou2019genuine,boreiri2023towards,MO_generic_2022,MO_rigidity_2022}, or at least that no straightforward reduction to a bipartite nonlocality test is known. These examples rely crucially on the interplay between the network topology and the use of entangled measurements.

Typically, whether a probability distribution admits a classical model is assessed through the violation of a Bell inequality. In the standard Bell scenario, the algorithms used to characterize which probability distributions are classically realizable are well understood. By contrast, deriving Bell-like inequalities in network scenarios is substantially more challenging, primarily due to the intrinsic nonconvexity of the corresponding classical sets. Although some numerical methods are available—most notably the inflation technique~\cite{WolfeSpekkensFritz_2019,wolfe2021quantum}, their computational complexity often limits their practical applicability. Here, to overcome these difficulties, we employ latent splitting to derive new, robust Bell-like inequalities for the triangle network. These inequalities allow us to certify the nonclassicality of the paradigmatic RGB4 family~\cite{renou2019genuine} of probability distributions in the triangle network with four outcomes.
 
The family of distributions studied in \cite{renou2019genuine} can be generated from measurements on bipartite quantum states distributed according to the triangle network (see Fig.~\ref{fig:examples-DAG}), and admits an interpretation in terms of excitation counting \cite{MO_rigidity_2022,MO_generic_2022}. Consider the three sources in the triangle network, $\rho_{\alpha}$, $\rho_{\beta}$, and $\rho_{\gamma}$, each distributing an entangled two-qubit state
\begin{equation}
\ket{\psi} = \lambda_0\ket{01} + \lambda_1\ket{10},
\end{equation}
with positive real coefficients $\lambda_{0}$ and $\lambda_{1}$ satisfying $\lambda_0^2 + \lambda_1^2 = 1$. These states can be interpreted as an excitation being sent either to the left ($\ket{10}$) or to the right ($\ket{01}$). In what follows, we focus on the case $\lambda_0=\sqrt{\frac{2}{3}}$.

Each party then measures the total number of excitations it receives. The states $\ket{\bar{0}}=\ket{00}$ and $\ket{\bar{2}}=\ket{11}$ correspond to receiving zero or two excitations, respectively, while single-excitation events are measured in a superposition basis defined by the projectors onto the states $\ket{\bar{1}+}=u\ket{01}+v\ket{10}$ and $\ket{\bar{1}-}=v\ket{01}-u\ket{10}$, with $0\leq u \leq 1$ and $v=\sqrt{1-u^2}$. When all parties perform this measurement, the resulting statistics define a family (RGB4 family) of observational distributions $P_{obs}^u(a,b,c)$.
\subsubsection{Robust Certification and Witnesses}
\begin{figure}[t]
\begin{minipage}[t]{0.48\columnwidth}
(a) Latent splitting is performed on the edge $\gamma \to A$.\vspace{0.8cm}
    \centering
    \begin{tikzpicture}
        \draw (120 - 30: .8cm) node[latent] (l1) {$\beta$};
        \draw (120 + 30: 1.6cm) node[var] (a) {$A$};
        \draw (240 - 30: .8cm) node[latent] (l2) {$\gamma$};
        \draw (240 + 30: 1.6cm) node[var] (b) {$B$};
        \draw (360 - 30: .8cm) node[latent] (l3) {$\alpha$};
        \draw (360 + 30: 1.6cm) node[var] (c) {$C$};

        \path[dir, forestGreen] (l1) edge (a) (l1) edge (c) (l2) edge (a) (l2) edge (b) (l3) edge (b) (l3) edge (c); 
        \path[dir, burgundy] (l2) edge (a);
    \end{tikzpicture}
\end{minipage}
\begin{minipage}[t]{0.48\columnwidth}
(b) Joint DAG of the obs. and int. experiment
    \centering
    \begin{tikzpicture}

        \foreach [count=\k] \l/\n/\a in {1/A/\beta, 2/B/\gamma, 3/C/\alpha} {
            \draw (\k*360/3 - 30: .8cm) node[latent] (l\k) {$\a$};
            \draw (\k*360/3 + 30: 1.6cm) node[var] (\n) {$\n$};
			\path[dir, forestGreen] (l\k) edge (\n);
        }
		\foreach \k/\l in {1/C, 2/A, 3/B}
			\path[dir,forestGreen] (l\k) edge (\l);

        \draw (120 + 0: 2.2cm) node[var] (ai) {$\hat A$};
        \draw (120 + 30: 2.6cm) node[var] (l2i) {$\hat \gamma$};
        \path[dir,forestGreen] (l1) edge (ai) (l2i) edge (ai);
    \end{tikzpicture}
\end{minipage}
\caption{\textbf{Latent splitting in the triangle scenario.} (a)~Shows the latent splitting performed by $A$, cutting effectively the influence from $\gamma$. (b)~The joint DAG resulting from the intervention. }
\label{fig: triangle-PI}
\end{figure}
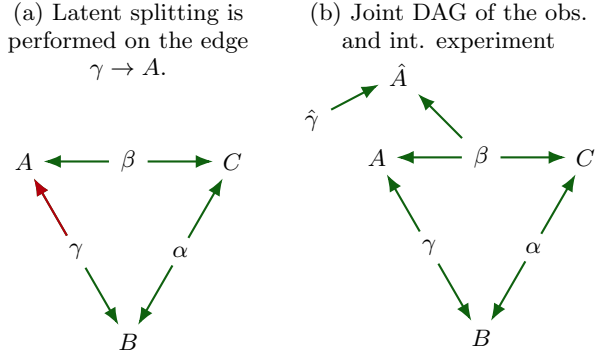

We now show how latent splitting allows for a robust certification of nonclassicality of the RGB4 distribution. Assume that the parties can implement latent splittings via local operations acting on their respective shares of the sources. In particular, we consider Alice performing a latent splitting on the source she shares with Bob, that is, on the edge $\gamma\rightarrow A$.
In Fig.~\ref{fig: triangle-PI}(a) we show the intervened edge. 

Operationally, this requires all parties to recollect their data while considering a new global state, given by $\rho_{int}=\rho_{\alpha}\otimes \rho_{\beta}\otimes (\text{Tr}_B\rho_\gamma\otimes\text{Tr}_A\rho_\gamma)$, which in turn gives rise to a corresponding family of interventional distributions $P^u_{int}(a,b,c)$. Our goal is to show that there cannot exist a single classical causal model—specified by the collection of probability distributions $p(\alpha)$, $p(\beta)$, $p(\gamma)$, $p(a|\beta,\gamma)$, $p(b|\alpha,\gamma)$, and $p(c|\alpha,\beta)$—that simultaneously satisfies
\begin{align}
    &\nonumber P^u_{obs}(a,b,c)=\sum_{\alpha,\beta,\gamma}p(\alpha)p(\beta)p(\gamma)p(a|\beta,\gamma)p(b|\alpha,\gamma)p(c|\alpha,\beta),\\
    &P^u_{int}(a,b,c)=\sum_{\alpha,\beta,\gamma}p(\alpha)p(\beta)p(\gamma)p(a|\beta)p(b|\alpha,\gamma)p(c|\alpha,\beta) ,\label{eq: Int_obs_cModel}
\end{align}
where $p(a|\beta):=\sum_{\hat{\gamma}}p(\hat{\gamma})p(a|\beta,\hat{\gamma})$.

%where $p(a|\beta):=\sum_{\gamma}p(\gamma)p(a|\beta,\gamma)$, and analogously for $p(b|\alpha)$. %While it is already known that $P^u_{obs}$ cannot be reproduced by any classical model, the additional constraints implied by the existence of the interventional distribution $P^u_{int}$ allow us to go one step further. In particular, they enable the derivation of new Bell-like inequalities for the triangle network using the inflation technique (see Appendix \ref{app: inflation_RGB4} for a more detailed derivation).

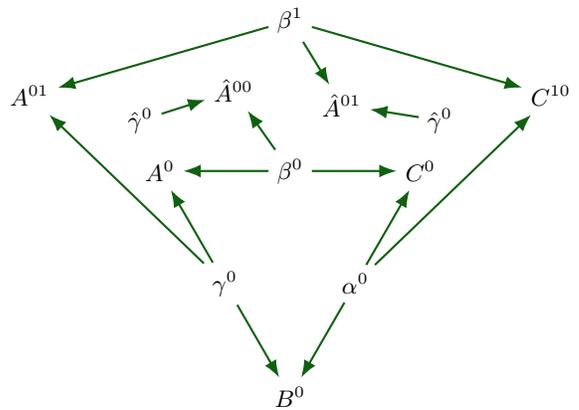
\begin{figure}[h]
    \centering
    \begin{tikzpicture}

        \foreach [count=\k] \l/\n/\a in {1/A^0/\beta^0, 2/B^0/\gamma^0, 3/C^0/\alpha^0} {
            \draw (\k*360/3 - 30: 1cm) node[latent] (l\k) {$\a$};
            \draw (\k*360/3 + 30: 2cm) node[var] (\n) {$\n$};
			\path[dir, forestGreen] (l\k) edge (\n);
        }
		\foreach \k/\l in {1/C^0, 2/A^0, 3/B^0}
			\path[dir, forestGreen] (l\k) edge (\l);

        \draw (120 - 10: 2.2cm) node[var] (ai) {$\hat A^{00}$};
        \draw (120 + 20: 2.6cm) node[var] (l2i) {$\hat \gamma^0$};
    
        \path[dir, forestGreen] (l1) edge (ai) (l2i) edge (ai);

        \draw (120 + 30: 4cm) node[var] (B1) {$A^{01}$};
        \draw (360 + 30: 4cm) node[var] (A1) {$C^{10}$};
        \draw (360/3 - 30: 3cm) node[latent] (beta1) {$\beta^1$};

        \draw (120 -50: 2.cm) node[var] (Ap1) {$\hat A^{01}$};
        \draw (120 -80: 2.6cm) node[var] (gamma0i) {$\hat\gamma^0$};

        \path[dir, forestGreen] (beta1) edge (B1)  (beta1) edge (A1) (beta1) edge (Ap1);
        \path[dir, forestGreen] (gamma0i) edge (Ap1);
        \path[dir, forestGreen] (l2) edge (B1)  ;
        \path[dir, forestGreen] (l3) edge (A1)  ;
    \end{tikzpicture}
\caption{\textbf{Inflation of the Joint DAG in the triangle scenario.} An inflation  of the Joint Causal DAG shown in Fig.~\ref{fig: triangle-PI}(b). The duplicated source is $\beta$, and since it is a classical latent variable it can distribute copies of share randomness to copies of $A$, $\hat A$ and $C$.}
\label{fig: inflation_RGB4}
\end{figure}

To show this, we assume the contrary. We assume that there exists a classical model such that reproduces both probabilities distributions. Then $\beta$ can be copied and distributed to a party $\hat A$, representing the after-intervention version of variable $A$. This new scenario has associated the causal structure shown in Fig.~\ref{fig: triangle-PI}(b). In this scenario,  there must exist a probability distribution $q(a,\hat a,b,c)$ satisfying
 
\begin{align}
    P^u_{obs}(a,b,c) = \sum_{\hat a}q(a,\hat a,b,c),  \nonumber\\   
    P^u_{int}(\hat a,b,c) = \sum_{ a}q(a,\hat a,b,c).\label{eq: conditions_g}
\end{align}
To assess the existence of such a distribution, we use the Python LP Inflation package provided in Ref.~\cite{pythoninflation}. This package imposes the usual inflation constraints (see Ref.~\cite{WolfeSpekkensFritz_2019}), and allows to manually add constraints~(\ref{eq: conditions_g}). Remarkably, by considering the inflation shown in Fig.~\ref{fig: inflation_RGB4}, which is obtained by only considering a single copy of $\beta$, one can already prove the nonexistence of $q$, showing the nonclassicality of the distribution.
We emphasize that with this same level of inflation, the LP can find a feasible solution when considering only the observational distribution, showing the relevant role of the intervention. 

By looking at the dual program one is able to extract a witness $\mathcal{I}$. For example, for the choice of parameter $u=0.85$,  after some algebraic manipulation one obtains the inequality 
\begin{align}
\mathcal{I} &= P^u_{obs}(a=3,c=\{1,2\})  - P^u_{int}(3,2,\{0,1\})  \nonumber\\
& - P^u_{int}(3,0,\{1,2\}) + P^u_{obs}(b=2) (P^u_{obs}(a=3))^2 \nonumber \\
&+P^u_{obs}(a=3)P^u_{obs}(\{0,2\},0,\{1,3\}) - 2P^u_{int}(3,2,2) \nonumber \\
& - P^u_{obs}(a=\{0,2\}) P^u_{obs}(a=3,c=\{1,2\}) \nonumber \\
& + (1- P^u_{obs}(a=3)) \Big[ P^u_{int}(3,2,\{0,1\})  + 2P^u_{int}(3,2,2) \Big]  \nonumber \\
&+P^u_{obs}(a=3)P^u_{obs}(\{0,2\},\{0,2\},\{0,2\}) \ge 0. \nonumber \\
\label{eq:simplified_witness}
\end{align}
The notation  $P^u_{int}(3,2,\{0,1\})$ indicates one should sum over the values of $c=0$ and $c=1$:
$P^u_{int}(3,2,\{0,1\}) = P^u_{int}(3,2,0) + P^u_{int}(3,2,1)$. When a marginal is present, we denote explicitly the parties through the label outcome, for example $P^u_{obs}(a=3,c=\{1,2\})$. After substituting $P^u_{obs}$ and $P^u_{int}$ in the left hand side of  Ineq.(\ref{eq:simplified_witness}), one obtains  $\mathcal{I} \approx -2.5\times10^{-4}< 0$ witnessing  the nonclassical behaviour.

\begin{figure}
    \centering
    \includegraphics{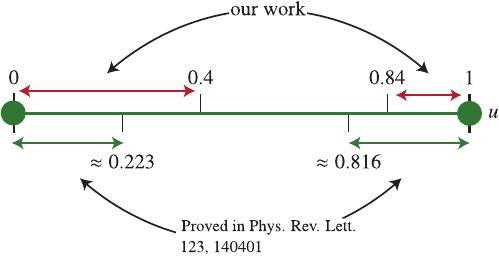}
    \caption{\textbf{The RGB4 distribution for distinct values of $u$.} Here we compare different intervals of certification corresponding to nonclassical correlations $P^u_{obs}$ and $P^u_{int}$. The green lower intervals, $u\in(0,0.223)$ and $u\in(0.816,1)$ indicate the result proved in \cite{renou2019genuine}, which exclude imperfect preparations or noisy measurements. Our work (red upper intervals) identifies first a broader region, $u\in(0,0.4)$ and, a slightly smaller region, $u\in(0.84,1)$, obtained through the inflation technique.}
    \label{fig: Infeasibility_u}
\end{figure}
In Fig. \ref{fig: Infeasibility_u}, we show certification intervals for the RGB4 distribution as a function of the measurement parameter $u$ and compare our methods with previous results proved by Renou et al.~\cite{renou2019genuine} obtained by the token counting technique. The different intervals indicate regions of $u$ such that the correlations obtained are certifiably nonclassical. While Ref.~\cite{renou2019genuine} certifies nonclassicality only for $u\in(0,0.223)$ and  $u\in(0.816,1)$, our method based on the inflation technique identifies an extended lower region  $u\in(0,0.4)$ and a slightly narrower upper one $u\in(0.84,1)$. As mentioned previously, if we consider only the observational distribution, we can find a feasible solution with the same level of inflation used in this work, showing a clear detection of nonclassicality improvement by considering latent splitting. Furthermore, by only employing a small inflation level, the framework remains highly relevant for practical implementations, as latent splitting offers additional power to derive network Bell-like inequalities that can be directly tested in realistic scenarios. This is in contrast to other techniques, such as token counting that although powerful,  it fails to consider distributions that are close to but outside the RGB4 family (such as those that may arise in experimental realizations). In what follows we explore the noise tolerance of the RGB4 distribution under the latent splitting framework.

For each source, we consider  depolarized states of the form 
\begin{equation}
    \rho_D= v\ket{\psi}\bra{\psi} + (1-v)\mathbb I /4.
\end{equation}
 To account for sources with different levels of noise we define the visibilities $v_\alpha,v_\beta$ and $v_\gamma $. First, we consider the symmetric case, when all sources are equal so $v_\alpha=v_\beta=v_\gamma = v $. In this case we obtain a minimum visibility of $v=0.9971$. In the asymmetric case, we can find lower visibilities and furthermore, the asymmetry on the causal structure (see Fig.~\ref{fig: triangle-PI}(b)) reflects on the minimum visibility of each individual source.
 
 For example, if $v_\beta = v_\gamma =1$ a minimum visibility of  $v_\alpha = 0.9946$ is needed for the distribution to be non-classical. On the other hand, when $v_\alpha = v_\beta =1$  the minimum visibility for the last source is $ v_\gamma = 0.988$. Finally, when considering $v_\alpha = v_\gamma  =1$, the minimum visibility for $\rho_\beta$ is $v_\beta = 0.9854$.  This hierarchy of critical visibilities ($v_\beta < v_\gamma < v_\alpha$) provides insight into the structural dependencies of the post-intervention network. It indicates that the non-classical correlations are most sensitive to noise in source $\alpha$, while exhibiting the greatest resilience to noise in source $\beta$. 

In summary, our derivation shows that our framework of latent splitting allows to detect non-classicality in quantum networks while obtaining analytical witnesses in the process. The witness $\mathcal{I}$, combined with the noise tolerance analysis, motivates the use of latent splitting as not only a theoretical feature but also as a robust enough tool that allows for proofs of nonclassicality in realistic experimental implementations with imperfect sources.

\subsection{Beyond quantum nonclassicality with latent splitting}
Here, we show how one can go beyond the standard notion of network nonclassicality using latent splitting. That is, we show the existence of statistics which would never admit a certification of network nonclassicality but, nevertheless, do not admit a classical causal model explanation when we consider all the available data from the experiments. In particular, we show here a quantum strategy in the triangle network such that its observed statistics admit a classical explanation, but when we consider additional data from latent splitting we can prove their incompatibility. The strategy is based on the ``Fritz trick" which we briefly review in the following.

The first strategy that was provably non-classical in the triangle network was the Fritz model \cite{fritz2012beyond}. The idea is to have Alice and Bob perform a standard CHSH Bell test. To achieve this, Alice and Bob share a maximally entangled Bell state $|\psi^+ \rangle=\frac{1}{\sqrt{2}}(|00\rangle+|11\rangle)$, and the inputs that are required to test the CHSH inequality are provided by the two additional sources. The sources $\alpha$ and $\beta$ provide a uniformly random bit $y = 0,1$ and $x= 0,1$ respectively.  Upon receiving their effective inputs $x$ and $y$, Alice performs the Pauli measurements $\sigma_z$ or $\sigma_x$ and Bob performs $(\sigma_z+\sigma_x)/\sqrt{2}$ or $(\sigma_z-\sigma_x)/\sqrt{2}$ obtaining binary outputs $a'$ and $b'$. Finally, Alice outputs $a = (a',x)$, Bob $b = (b',y)$, and Charlie $c = (x,y)$. 

Fritz proved that since the values of the outputs $x(y)$ are perfectly correlated between Charlie and Alice (Charlie and Bob), these outputs must be independent of the source connecting Alice and Bob~\cite{fritz2012beyond}. Note that it is crucial that Charlie also outputs the values of $x$ and $y$ to ensure this independence. The resulting distribution $P_F(a,b,c)$ is nonlocal whenever the conditional distribution $P(a',b'|x,y)$ violates the CHSH Bell inequality. Notably, this argument is not limited to the CHSH inequality and can be applied to any bipartite Bell inequality.

This strategy yields a distribution with 4 outcomes for each party, and when adding noise to the Bell state shared by Alice and Bob, one finds a critical visibility of $v=\frac{1}{\sqrt{2}}$. However, even though 4 outcomes are sufficient to show nonclassicality in the triangle network, the authors in \cite{boreiri2023towards} showed that coarse-grainings of the Fritz distribution $P_F$ can also be certifiably nonclassical. Using the inflation technique, they show that the resulting distribution, with cardinality 3-3-2, is nonlocal for visibility greater than $v\approx 0.87$, higher compared to the critical visibility of the original Fritz distribution. Using numerical methods based on machine learning, they find that the distribution still has a critical visibility close to the original value of $\frac{1}{\sqrt{2}}\approx 0.71$. Additionally, they show that further coarse-graining of the distribution, to binary outcomes, can be classically reproduced with excellent accuracy. Here, we show that latent splitting can help us go beyond the standard certification of network nonclassicality, i.e. even though the observational distribution of the Fritz strategy can be reproduced by a classical model when coarse-grained to binary outcomes, this is not true anymore if we consider latent splitting. By incorporating this extra data, we can certify the nonclassicality of this distribution using the inflation technique.

In particular, we consider a distribution $P_{obs}(a,b,c)$ in the binary-outcome configuration of the triangle network and, additionally, we assume that Alice and Bob can perform a latent splitting on the sources $\beta$ and $\alpha$, respectively. Then we obtain the interventional distributions $P^{\beta}_{int}(a,b,c)$,  $P^{\alpha}_{int}(a,b,c)$, and  $P^{\alpha\beta}_{int}(a,b,c)$ representing the cases when we suppress the edge $\beta \to A$, the edge $\alpha \to B$, or both. Here, our proof uses the inflation technique: we start assuming that there exists a classical causal model for our strategy, i.e. there exists causal parameters $p(\alpha)$, $p(\beta)$,  $p(\gamma)$, and $p(a|\beta\gamma)$, $p(b|\alpha\gamma)$,  $p(c|\alpha\beta)$ s.t. it recovers all data tables. Consequently, we can distribute the classical information of $\gamma$ to other parties and we consider a new version of $\gamma$ that sends information to $\hat A$ and $\hat B$, which correspond to the variables that are probed with interventions; this scenario is shown in Fig.~\ref{fig: triangle_2pint}. 
In this new scenario, there must exist a global probability distribution $q(a,\hat a, b, \hat b, c)$ which satisfies

\begin{align}
    &P_{obs}(a,b,c) = \sum_{\hat a, \hat b} q(a,\hat a,b,\hat b, c), \nonumber \\
    &P^{\alpha}_{int}(a,\hat b,c), =  \sum_{\hat a, b} q(a,\hat a,b,\hat b, c),  \nonumber\\
    &P^{\beta}_{int} (\hat a,b,c) = \sum_{a, \hat b} q(a,\hat a,b,\hat b, c),\nonumber \\
    &P^{\alpha\beta}_{int}(\hat a,\hat b,c) = \sum_{a, b} q(a,\hat a,b,\hat b, c).
\end{align}
 Additionally, since we are only removing the causal connection without changing any other mechanism, we also assume that the latent variables $\hat \alpha, \hat \beta$ have the same distribution as $\alpha$ and $\beta$. This condition constraints the marginals $P_{obs}(a,b), P^{\beta}_{int} (\hat a,b), P^{\alpha}_{int}(a,\hat b)$ and $P^{\alpha\beta}_{int}(\hat a,\hat b)$ to be all equal. Now, defining the variables
 
 \begin{align*}
     &E_{obs}^{c}:=P_{obs}(a=b,c)-P_{obs}(a\neq b,c),\\
     &E_{\beta}^{c}:=P^{\beta}_{int} (\hat a=b,c)-P^{\beta}_{int} (\hat a\neq b,c),\\
     &E_{\alpha}^{c}:=P^{\alpha}_{int} (a=\hat b,c)-P^{\alpha}_{int} (a\neq \hat b,c),\\
     &E_{\alpha\beta }:=P^{\alpha \beta}_{int} (\hat a=\hat b)-P^{\alpha \beta}_{int} (\hat a\neq \hat b),
 \end{align*}
we are able to prove the following inequality :
 \begin{equation}
    S\equiv E_{\alpha\beta}P_{obs}(c=1)+2P_{obs}(c=1) - E^1_{obs}-E^1_{\alpha}-E^1_{\beta} \geq 0.
    \label{eq:chsh_2pint}
\end{equation}
 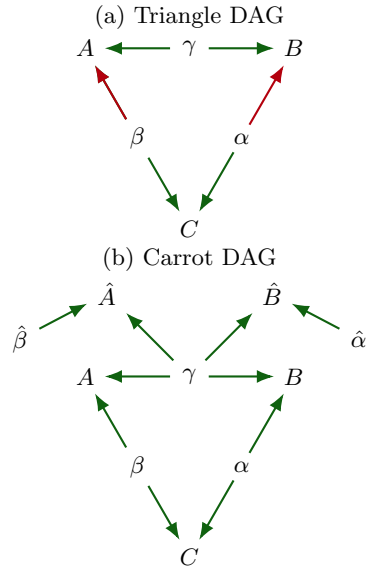
\begin{figure}[h]
\begin{minipage}[t]{1\columnwidth}
(a) Triangle DAG
    \centering
    
    \begin{tikzpicture}
        \draw (120 - 30: .8cm) node[latent] (l1) {$\gamma$};
        \draw (120 + 30: 1.6cm) node[var] (a) {$A$};
        \draw (240 - 30: .8cm) node[latent] (l2) {$\beta$};
        \draw (240 + 30: 1.6cm) node[var] (c) {$C$};
        \draw (360 - 30: .8cm) node[latent] (l3) {$\alpha$};
        \draw (360 + 30: 1.6cm) node[var] (b) {$B$};

        \path[dir][forestGreen] (l1) edge (a) (l1) edge (b) (l2) edge (c) (l2) edge (a) (l3) edge (c); 
        \path[dir, burgundy] (l2) edge (a);
        \path[dir, burgundy] (l3) edge (b);
    \end{tikzpicture}

\end{minipage}
\begin{minipage}[t]{1\columnwidth}
(b) Carrot DAG
    \centering
    
    \begin{tikzpicture}

        \foreach [count=\k] \l/\n/\a in {1/A/\gamma, 2/C/\beta, 3/B/\alpha} {
            \draw (\k*360/3 - 30: .8cm) node[latent] (l\k) {$\a$};
            \draw (\k*360/3 + 30: 1.6cm) node[var] (\n) {$\n$};
			\path[dir, forestGreen] (l\k) edge (\n);
        }
		\foreach \k/\l in {1/B, 2/A, 3/C}
			\path[dir, forestGreen] (l\k) edge (\l);

        \draw (120 + 0: 2.2cm) node[var] (ai) {$\hat A$};
        \draw (360 + 60: 2.2cm) node[var] (bi) {$\hat B$};
        \draw (120 + 30: 2.6cm) node[var] (l2i) {$\hat \beta$};
        \draw (360 + 30: 2.6cm) node[var] (l3i) {$\hat \alpha$};
        \path[dir, forestGreen] (l1) edge (ai) (l2i) edge (ai);
        \path[dir, forestGreen] (l1) edge (bi) (l3i) edge (bi);
    \end{tikzpicture}

\end{minipage}
\caption{\textbf{Latent splittings in the triangle network.} In a), we show the DAG of the triangle network and highlight the edges corresponding to a latent splitting on $\beta \to A$ and $\alpha \to B$. In b), we show the inflated DAG for the classically-equivalent joint scenario.}
\label{fig: triangle_2pint}
\end{figure}

 This inequality and the strategy that violates it can be understood as a CHSH game that uses the intervened edges and the value $c$ of Charlie to define settings, and, thus, it can be seen as the interventional version of the original ``Fritz trick" in the triangle network. In particular, we consider that the sources maintain their distributions and that we are not free to modulate the distributions of the  $\alpha$ and $\beta$ sources.

We will now show that we can violate~\eqref{eq:chsh_2pint} with the following quantum strategy: $\gamma$ prepares a maximally entangled state $|\psi^+ \rangle$ for Alice and Bob, $\alpha$ and $\beta$ distribute classical bits, with $ p(\alpha = 1) = p(\beta = 1) = \epsilon$, and Alice and Bob will perform the standard measurements for the CHSH violation based on $\alpha$ and $\beta$. However, Alice outputs $a$, Bob outputs $b$, and Charlie outputs $c=\alpha \beta$, i.e. the logical \textit{and} of $\alpha$ and $\beta$. Using this configuration, we can easily compute the value of~\eqref{eq:chsh_2pint}:
\begin{equation}
  S _Q  = \frac{4}{\sqrt{2}}( \epsilon-1)\epsilon^2-\frac{2}{\sqrt{2}} \epsilon^4 +2\epsilon^2,\label{eq: Interventional_fritz} 
\end{equation}
and we achieve $S_Q<0$ for $\epsilon < 1-\sqrt{\sqrt{2} - 1} \approx 0.3564$ (see Appendix \ref{app: Fritz_trick} for a detailed derivation). If we consider instead a noisy state, $\rho_\gamma = v \ket{\psi^-}\bra{\psi^-} +(1-v) I/2$, the values of $\epsilon$ leading to a violation are $\epsilon < 1-\sqrt{\sqrt{2}/v - 1}$, which in turn means that to be able to see a violation for a fixed $\epsilon$ we need a visibility of $v_{\mathrm{min}} = \frac{\sqrt{2} }{(1 + (1-\epsilon)^2)}$.

\section{Final remarks}\label{sec: conclusion}
\label{sec:conclusions}

In this work, we introduced and formalized the framework of latent splitting, a new causal tool tailored to quantum networks. Motivated by the operational ability to prepare, transform, and control quantum systems, latent splitting extends Pearl’s notion of intervention beyond observed variables to the latent quantum resources that mediate correlations. As a consequence of our new interventional probing scheme, we are able to incorporate a class of non-fanout inflations as experimentally accessible data that can be used to increment the analysis of quantum non-classicality. Furthermore, we showed that this framework strictly generalizes standard Pearl-like interventions: the classical do-calculus emerges as a special case, providing a unified language for classical and quantum causal manipulations.

The advantages of this generalization are particularly pronounced in network scenarios where standard interventions are either trivial or uninformative. Focusing on the triangle network, we demonstrated that latent splitting unlocks forms of non-classicality that remain hidden at the purely observational level. For the RGB4 family of distributions \cite{renou2019genuine}, we combined latent splitting with the inflation technique to derive polynomial inequalities witnessing genuine network nonlocality, extending the range of detectable nonclassical correlations and establishing robustness to noise. Moreover, we showed that latent splitting enables the certification of nonclassicality in the minimal triangle scenario with binary variables, by explicitly constructing a nonlinear Bell inequality. This result sheds light to a longstanding open question concerning the classical reproducibility of the minimal Fritz distribution.

Beyond these concrete results, latent splitting provides a conceptually new perspective on causality in quantum networks. By allowing controlled manipulations of latent quantum systems, they bridge causal inference and operational quantum theory more closely than is possible with observational data or standard interventions alone. Looking ahead, this framework can be applied to more general network topologies, such as networks containing intermediate latent variables \cite{centeno2025significance}, potentially revealing new forms of nonclassicality and strengthening device-independent certification protocols for quantum information processing tasks such as randomness generation and quantum key distribution in networked settings.
\\
\section{Acknowledgments}

 We thank Rafael Rabelo and Elie Wolfe for fruitful discussions. Research at Perimeter Institute is supported by the Government of Canada through the Department of Innovation, Science and Economic Development Canada and by the Province of Ontario through the Ministry of Research, Innovation and Science. Research at the International Institute of Physics was supported by the Simons Foundation (Grant No. 1023171, R.C.), the Brazilian National Council for Scientific and Technological Development (CNPq, Grants No. 307295/2020-6, No.403181/2024-0, and 301687/2025-0), a guest professorship from the Otto M\o nsted Foundation and the Financiadora de Estudos e Projetos (Grant No. 1699/24 IIF-FINEP), and the Coordenação de Aperfeiçoamento de Pessoal de Nível Superior, Brasil (CAPES) Finance Code 001. Research at Inria and École Polytechnique is supported by funding from
INRIA and CIEDS through the Action Exploratoire project DEPARTURE.
\bibliographystyle{apsrev4-2}
\bibliography{refs}

@article{Bikak_2019, 
title={Structural causal models’ application in development of clinical trials}, 
volume={7},
url={https://pulmonarychronicles.com/index.php/pulmonarychronicles/article/view/587}, 
DOI={10.12746/swrccc.v7i31.587}, 
number={31}, 
journal={The Southwest Respiratory and Critical Care Chronicles}, 
author={Bikak, Marvi}, 
year={2019}, 
month={Nov.}, 
pages={1-2} }

@article{Gani2023,
  author    = {Gani, Md Osman and Kethireddy, Shravan and Adib, Riddhiman and Hasan, Uzma and Griffin, Paul M. and Adibuzzaman, Mohammad},
  title     = {Structural causal model with expert augmented knowledge to estimate the effect of oxygen therapy on mortality in the ICU},
  journal   = {Artificial Intelligence in Medicine},
  year      = {2023},
  volume    = {137},
  pages     = {102493},
  doi       = {10.1016/j.artmed.2023.102493}
}

@article{MO_rigidity_2022,
  title = {Network nonlocality via rigidity of token counting and color matching},
  author = {Renou, Marc-Olivier and Beigi, Salman},
  journal = {Physical Review A},
  volume = {105},
  issue = {2},
  pages = {022408},
  numpages = {21},
  year = {2022},
  month = {Feb},
  publisher = {American Physical Society},
  doi = {10.1103/PhysRevA.105.022408},
  url = {https://link.aps.org/doi/10.1103/PhysRevA.105.022408}
}

@article{MO_generic_2022,
  title = {Nonlocality for Generic Networks},
  author = {Renou, Marc-Olivier and Beigi, Salman},
  journal = {Physical Review Letters},
  volume = {128},
  issue = {6},
  pages = {060401},
  numpages = {6},
  year = {2022},
  month = {Feb},
  publisher = {American Physical Society},
  doi = {10.1103/PhysRevLett.128.060401},
  url = {https://link.aps.org/doi/10.1103/PhysRevLett.128.060401}
}

@article{gitton2025,
      title={The Elegant Joint Measurement is Non-Classical in the Triangle Network}, 
      author={Victor Gitton and Renato Renner},
      year={2025},
      eprint={2510.15143},
      journal={arXiv preprint arXiv:2510.15143},
      primaryClass={quant-ph},
      url={https://arxiv.org/abs/2510.15143}, 
}

@article{Abiuso2022,
  title = {Single-photon nonlocality in quantum networks},
  author = {Abiuso, Paolo and Kriv\'achy, Tam\'as and Boghiu, Emanuel-Cristian and Renou, Marc-Olivier and Pozas-Kerstjens, Alejandro and Ac\'{\i}n, Antonio},
  journal = {Physical Review Research},
  volume = {4},
  issue = {1},
  pages = {L012041},
  numpages = {6},
  year = {2022},
  month = {Mar},
  publisher = {American Physical Society},
  doi = {10.1103/PhysRevResearch.4.L012041},
  url = {https://link.aps.org/doi/10.1103/PhysRevResearch.4.L012041}
}

@article{Renou2022,
  title = {Nonlocality for Generic Networks},
  author = {Renou, Marc-Olivier and Beigi, Salman},
  journal = {Physical Review Letters},
  volume = {128},
  issue = {6},
  pages = {060401},
  numpages = {6},
  year = {2022},
  month = {Feb},
  publisher = {American Physical Society},
  doi = {10.1103/PhysRevLett.128.060401},
  url = {https://link.aps.org/doi/10.1103/PhysRevLett.128.060401}
}

@article{Branciard2012,
  title = {Bilocal versus nonbilocal correlations in entanglement-swapping experiments},
  author = {Branciard, Cyril and Rosset, Denis and Gisin, Nicolas and Pironio, Stefano},
  journal = {Physical Review A},
  volume = {85},
  issue = {3},
  pages = {032119},
  numpages = {21},
  year = {2012},
  month = {Mar},
  publisher = {American Physical Society},
  doi = {10.1103/PhysRevA.85.032119},
  url = {https://link.aps.org/doi/10.1103/PhysRevA.85.032119}
}

@book{serafini2017quantum,
  title={Quantum Continuous Variables: A Primer of Theoretical Methods},
  author={Serafini, A.},
  isbn={9781482246346},
  lccn={2016058596},
  url={https://books.google.dk/books?id=zHtgvgAACAAJ},
  year={2017},
  publisher={CRC Press, Taylor \& Francis Group}
}

@book{Nielsen_Chuang_2010, 
    place={Cambridge}, 
    title={Quantum Computation and Quantum Information: 10th Anniversary Edition}, publisher={Cambridge University Press}, author={Nielsen, Michael A. and Chuang, Isaac L.}, year={2010},
    url = {https://www.cambridge.org/highereducation/books/quantum-computation-and-quantum-information/01E10196D0A682A6AEFFEA52D53BE9AE?utm_campaign=shareaholic&utm_medium=copy_link&utm_source=bookmark}
    }

@article{pythoninflation,
  title={Inflation: a {P}ython library for classical and quantum causal compatibility},
  author={Boghiu, Emanuel-Cristian and Wolfe, Elie and Pozas-Kerstjens, Alejandro},
  journal={Quantum},
  url = {https://doi.org/10.22331/q-2023-05-04-996},
  volume={7},
  pages={996},
  year={2023},
  publisher={{Verein zur F{\"{o}}rderung des Open Access Publizierens in den Quantenwissenschaften}}
}

@article{WolfeSpekkensFritz_2019,
  title={The Inflation Technique for Causal Inference with Latent Variables},
  author={Elie Wolfe and Robert W. Spekkens and Tobias Fritz},
  journal={Journal of Causal Inference},
  url = {https://doi.org/10.1515/jci-2017-0020},
  volume={7},
  number={2},
  pages={20170020},
  year={2019},
  publisher={De Gruyter}
}

@article{brukner2004bell,
  title={Bell’s inequalities and quantum communication complexity},
  author={Brukner, {\v{C}}aslav and {\.Z}ukowski, Marek and Pan, Jian-Wei and Zeilinger, Anton},
  journal={Physical review letters},
  url = {https://doi.org/10.1103/PhysRevLett.92.127901},
  volume={92},
  number={12},
  pages={127901},
  year={2004},
  publisher={APS}
}

@article{buhrman2010nonlocality,
  title={Nonlocality and communication complexity},
  author={Buhrman, Harry and Cleve, Richard and Massar, Serge and De Wolf, Ronald},
  journal={Reviews of modern physics},
  url = {https://doi.org/10.1103/RevModPhys.82.665},
  volume={82},
  number={1},
  pages={665--698},
  year={2010},
  publisher={APS}
}

@article{wolfe2021quantum,
  title={Quantum inflation: A general approach to quantum causal compatibility},
  author={Wolfe, Elie and Pozas-Kerstjens, Alejandro and Grinberg, Matan and Rosset, Denis and Ac{\'\i}n, Antonio and Navascu{\'e}s, Miguel},
  journal={Physical Review X},
  url = {https://doi.org/10.1103/PhysRevX.11.021043},
  volume={11},
  number={2},
  pages={021043},
  year={2021},
  publisher={APS}
}

@article{gu2023experimental,
  title={Experimental full network nonlocality with independent sources and strict locality constraints},
  author={Gu, Xue-Mei and Huang, Liang and Pozas-Kerstjens, Alejandro and Jiang, Yang-Fan and Wu, Dian and Bai, Bing and Sun, Qi-Chao and Chen, Ming-Cheng and Zhang, Jun and Yu, Sixia and others},
  journal={Physical Review Letters},
  url = {https://doi.org/10.1103/PhysRevLett.130.190201},
  volume={130},
  number={19},
  pages={190201},
  year={2023},
  publisher={APS}
}

@article{fritz2012beyond,
  title={Beyond Bell's theorem: correlation scenarios},
  author={Fritz, Tobias},
  journal={New Journal of Physics},
  url = {https://doi.org/10.1088/1367-2630/14/10/103001},
  volume={14},
  number={10},
  pages={103001},
  year={2012},
  publisher={IOP Publishing}
}

@article{renou2019genuine,
  title={Genuine quantum nonlocality in the triangle network},
  author={Renou, Marc-Olivier and B{\"a}umer, Elisa and Boreiri, Sadra and Brunner, Nicolas and Gisin, Nicolas and Beigi, Salman},
  journal={Physical review letters},
  url = {https://doi.org/10.1103/PhysRevLett.123.140401},
  volume={123},
  number={14},
  pages={140401},
  year={2019},
  publisher={APS}
}

@article{chaves2021causal,
  title={Causal networks and freedom of choice in bell’s theorem},
  author={Chaves, Rafael and Moreno, George and Polino, Emanuele and Poderini, Davide and Agresti, Iris and Suprano, Alessia and Barros, Mariana R and Carvacho, Gonzalo and Wolfe, Elie and Canabarro, Askery and others},
  journal={PRX Quantum},
  url = {https://doi.org/10.1103/PRXQuantum.2.040323},
  volume={2},
  number={4},
  pages={040323},
  year={2021},
  publisher={APS}
}

@article{weilenmann2020self,
  title={Self-testing of physical theories, or, is quantum theory optimal with respect to some information-processing task?},
  author={Weilenmann, Mirjam and Colbeck, Roger},
  journal={Physical Review Letters},
  url = {https://doi.org/10.1103/PhysRevLett.125.060406},
  volume={125},
  number={6},
  pages={060406},
  year={2020},
  publisher={APS}
}

@article{suprano2022experimental,
  title={Experimental genuine tripartite nonlocality in a quantum triangle network},
  author={Suprano, Alessia and Poderini, Davide and Polino, Emanuele and Agresti, Iris and Carvacho, Gonzalo and Canabarro, Askery and Wolfe, Elie and Chaves, Rafael and Sciarrino, Fabio},
  journal={PRX Quantum},
  url = {https://doi.org/10.1103/PRXQuantum.3.030342},
  volume={3},
  number={3},
  pages={030342},
  year={2022},
  publisher={APS}
}

@article{pozas2022full,
  title={Full network nonlocality},
  author={Pozas-Kerstjens, Alejandro and Gisin, Nicolas and Tavakoli, Armin},
  journal={Physical review letters},
  url = {https://doi.org/10.1103/PhysRevLett.128.010403},
  volume={128},
  number={1},
  pages={010403},
  year={2022},
  publisher={APS}
}

@InProceedings{korb2004,
  title={Varieties of Causal Intervention},
  author={Korb, Kevin B. and Hope, Lucas R. and Nicholson, Ann E. and Axnick, Karl},
  editor={Zhang, Chengqi and W. Guesgen, Hans and Yeap, Wai-Kiang},
  booktitle={PRICAI 2004: Trends in Artificial Intelligence},
  url = {https://link.springer.com/chapter/10.1007/978-3-540-28633-2_35},
  pages={322--331},
  year={2004},
  publisher={Springer Berlin Heidelberg}
}

@article{lauand2024quantum,
  title={Quantum non-classicality in the simplest causal network},
  author={Lauand, Pedro and Poderini, Davide and Rabelo, Rafael and Chaves, Rafael},
  journal={arXiv preprint arXiv:2404.12790},
  url = {https://arxiv.org/abs/2404.12790},
  year={2024}
}

@book{pearl2009causality,
  title={Causality},
  author={Pearl, Judea},
  year={2009},
  publisher={Cambridge university press},
  url = {https://doi.org/10.1017/CBO9780511803161}
}

@article{lauand2024quantum2,
  title={Quantum Non-classicality from Causal Data Fusion},
  author={Lauand, Pedro and Bekele, Bereket Ngussie and Wolfe, Elie},
  journal={arXiv preprint arXiv:2405.19252},
  url = {https://arxiv.org/abs/2405.19252},
  year={2024}
}

@article{lauand2023witnessing,
  title={Witnessing nonclassicality in a causal structure with three observable variables},
  author={Lauand, Pedro and Poderini, Davide and Nery, Ranieri and Moreno, George and Pollyceno, Lucas and Rabelo, Rafael and Chaves, Rafael},
  journal={PRX Quantum},
  url = {https://doi.org/10.1103/PRXQuantum.4.020311},
  volume={4},
  number={2},
  pages={020311},
  year={2023},
  publisher={APS}
}

@article{poderini2024observational,
  title = {Observational-interventional Bell inequalities},
  author = {Poderini, Davide and Nery, Ranieri and Moreno, George and Zamora, Santiago and Lauand, Pedro and Chaves, Rafael},
  journal = {Physical Review A},
  volume = {110},
  issue = {4},
  pages = {042213},
  numpages = {14},
  year = {2024},
  month = {Oct},
  publisher = {American Physical Society},
  doi = {10.1103/PhysRevA.110.042213},
  url = {https://link.aps.org/doi/10.1103/PhysRevA.110.042213}
}

@article{gachechiladze2020quantifying,
  title={Quantifying causal influences in the presence of a quantum common cause},
  author={Gachechiladze, Mariami and Miklin, Nikolai and Chaves, Rafael},
  journal={Physical Review Letters},
  url = {https://doi.org/10.1103/PhysRevLett.125.230401},
  volume={125},
  number={23},
  pages={230401},
  year={2020},
  publisher={APS}
}

@article{agresti2022experimental,
  title={Experimental test of quantum causal influences},
  author={Agresti, Iris and Poderini, Davide and Polacchi, Beatrice and Miklin, Nikolai and Gachechiladze, Mariami and Suprano, Alessia and Polino, Emanuele and Milani, Giorgio and Carvacho, Gonzalo and Chaves, Rafael and others},
  journal={Science advances},
  url = {https://doi.org/10.1126/sciadv.abm1515},
  volume={8},
  number={8},
  pages={eabm1515},
  year={2022},
  publisher={American Association for the Advancement of Science}
}

@article{renou2021quantum,
  title={Quantum theory based on real numbers can be experimentally falsified},
  author={Renou, Marc-Olivier and Trillo, David and Weilenmann, Mirjam and Le, Thinh P and Tavakoli, Armin and Gisin, Nicolas and Ac{\'\i}n, Antonio and Navascu{\'e}s, Miguel},
  journal={Nature},
  url = {https://doi.org/10.1038/s41586-021-04160-4},
  volume={600},
  number={7890},
  pages={625--629},
  year={2021},
  publisher={Nature Publishing Group UK London}
}

@article{wehner2018quantum,
  title={Quantum internet: A vision for the road ahead},
  author={Wehner, Stephanie and Elkouss, David and Hanson, Ronald},
  journal={Science},
  url = {https://doi.org/10.1126/science.aam9288},
  volume={362},
  number={6412},
  pages={eaam9288},
  year={2018},
  publisher={American Association for the Advancement of Science}
}

@article{chen2021integrated,
  title={An integrated space-to-ground quantum communication network over 4,600 kilometres},
  author={Chen, Yu-Ao and Zhang, Qiang and Chen, Teng-Yun and Cai, Wen-Qi and Liao, Sheng-Kai and Zhang, Jun and Chen, Kai and Yin, Juan and Ren, Ji-Gang and Chen, Zhu and others},
  journal={Nature},
  url = {https://www.nature.com/articles/s41586-020-03093-8},
  volume={589},
  number={7841},
  pages={214--219},
  year={2021},
  publisher={Nature Publishing Group UK London}
}

@article{dynes2019cambridge,
  title={Cambridge quantum network},
  author={Dynes, JF and Wonfor, Adrian and Tam, WW-S and Sharpe, AW and Takahashi, R and Lucamarini, M and Plews, A and Yuan, ZL and Dixon, AR and Cho, J and others},
  journal={npj Quantum Information},
  url = {https://doi.org/10.1038/s41534-019-0221-4},
  volume={5},
  number={1},
  pages={101},
  year={2019},
  publisher={Nature Publishing Group UK London}
}

@article{simon2017towards,
  title={Towards a global quantum network},
  author={Simon, Christoph},
  journal={Nature Photonics},
  url = {https://doi.org/10.1038/s41566-017-0032-0},
  volume={11},
  number={11},
  pages={678--680},
  year={2017},
  publisher={Nature Publishing Group UK London}
}

@article{hutter2023quantifying,
  title={Quantifying quantum causal influences},
  author={Hutter, Lucas and Chaves, Rafael and Nery, Ranieri Vieira and Moreno, George and Brod, Daniel Jost},
  journal={Physical Review A},
  volume={108},
  number={2},
  pages={022222},
  year={2023},
  publisher={APS}
}

@article{chaves2018quantum,
  title={Quantum violation of an instrumental test},
  author={Chaves, Rafael and Carvacho, Gonzalo and Agresti, Iris and Di Giulio, Valerio and Aolita, Leandro and Giacomini, Sandro and Sciarrino, Fabio},
  journal={Nature Physics},
  volume={14},
  number={3},
  pages={291--296},
  year={2018},
  publisher={Nature Publishing Group UK London}
}

@article{Abergsemi2020,
  title = {Semidefinite Tests for Quantum Network Topologies},
  author = {\AA{}berg, Johan and Nery, Ranieri and Duarte, Cristhiano and Chaves, Rafael},
  journal = {Phys. Rev. Lett.},
  volume = {125},
  issue = {11},
  pages = {110505},
  numpages = {6},
  year = {2020},
  month = {Sep},
  publisher = {American Physical Society},
  doi = {10.1103/PhysRevLett.125.110505},
  url = {https://link.aps.org/doi/10.1103/PhysRevLett.125.110505}
}

@article{da2025local,
  title={Local models and Bell inequalities for the minimal triangle network},
  author={da Silva, Jos{\'e} M{\'a}rio and Pozas-Kerstjens, Alejandro and Parisio, Fernando},
  journal={Physical Review A},
  volume={112},
  number={3},
  pages={L030403},
  year={2025},
  publisher={APS}
}

@article{polino2023experimental,
  title={Experimental nonclassicality in a causal network without assuming freedom of choice},
  author={Polino, Emanuele and Poderini, Davide and Rodari, Giovanni and Agresti, Iris and Suprano, Alessia and Carvacho, Gonzalo and Wolfe, Elie and Canabarro, Askery and Moreno, George and Milani, Giorgio and others},
  journal={Nature Communications},
  url = {https://www.nature.com/articles/s41467-023-36428-w},
  volume={14},
  number={1},
  pages={909},
  year={2023},
  publisher={Nature Publishing Group UK London}
}

@article{saunders2017experimental,
  title={Experimental demonstration of nonbilocal quantum correlations},
  author={Saunders, Dylan J and Bennet, Adam J and Branciard, Cyril and Pryde, Geoff J},
  journal={Science advances},
  url = {https://doi.org/10.1126/sciadv.1602743},
  volume={3},
  number={4},
  pages={e1602743},
  year={2017},
  publisher={American Association for the Advancement of Science}
}

@article{poderini2020experimental,
  title={Experimental violation of n-locality in a star quantum network},
  author={Poderini, Davide and Agresti, Iris and Marchese, Guglielmo and Polino, Emanuele and Giordani, Taira and Suprano, Alessia and Valeri, Mauro and Milani, Giorgio and Spagnolo, Nicol{\`o} and Carvacho, Gonzalo and others},
  journal={Nature communications},
  url = {https://www.nature.com/articles/s41467-020-16189-6},
  volume={11},
  number={1},
  pages={2467},
  year={2020},
  publisher={Nature Publishing Group UK London}
}

@article{wang2024experimental,
  title={Experimental genuine quantum nonlocality in the triangle network},
  author={Wang, Ning-Ning and Zhang, Chao and Cao, Huan and Xu, Kai and Liu, Bi-Heng and Huang, Yun-Feng and Li, Chuan-Feng and Guo, Guang-Can and Gisin, Nicolas and Kriv{\'a}chy, Tam{\'a}s and others},
  journal={arXiv preprint arXiv:2401.15428},
  url = {https://arxiv.org/abs/2401.15428},
  year={2024}
}

@article{carvacho2017experimental,
  title={Experimental violation of local causality in a quantum network},
  author={Carvacho, Gonzalo and Andreoli, Francesco and Santodonato, Luca and Bentivegna, Marco and Chaves, Rafael and Sciarrino, Fabio},
  journal={Nature communications},
  url = {https://doi.org/10.1038/ncomms14775},
  volume={8},
  number={1},
  pages={14775},
  year={2017},
  publisher={Nature Publishing Group UK London}
}

@article{tavakoli2022bell,
  title={Bell nonlocality in networks},
  author={Tavakoli, Armin and Pozas-Kerstjens, Alejandro and Luo, Ming-Xing and Renou, Marc-Olivier},
  journal={Reports on Progress in Physics},
  url = {https://iopscience.iop.org/article/10.1088/1361-6633/ac41bb},
  volume={85},
  number={5},
  pages={056001},
  year={2022},
  publisher={IOP Publishing}
}

@article{komar2014quantum,
  title={A quantum network of clocks},
  author={Komar, Peter and Kessler, Eric M and Bishof, Michael and Jiang, Liang and S{\o}rensen, Anders S and Ye, Jun and Lukin, Mikhail D},
  journal={Nature Physics},
  url = {https://www.nature.com/articles/nphys3000},
  volume={10},
  number={8},
  pages={582--587},
  year={2014},
  publisher={Nature Publishing Group UK London}
}

@article{wei2019experimental,
  title={Experimental quantum switching for exponentially superior quantum communication complexity},
  author={Wei, Kejin and Tischler, Nora and Zhao, Si-Ran and Li, Yu-Huai and Arrazola, Juan Miguel and Liu, Yang and Zhang, Weijun and Li, Hao and You, Lixing and Wang, Zhen and others},
  journal={Physical review letters},
  url = {https://doi.org/10.1103/PhysRevLett.122.120504},
  volume={122},
  number={12},
  pages={120504},
  year={2019},
  publisher={APS}
}

@article{brunner2014bell,
  title={Bell nonlocality},
  author={Brunner, Nicolas and Cavalcanti, Daniel and Pironio, Stefano and Scarani, Valerio and Wehner, Stephanie},
  journal={Reviews of modern physics},
  url = {https://doi.org/10.1103/RevModPhys.86.419},
  volume={86},
  number={2},
  pages={419--478},
  year={2014},
  publisher={APS}
}

@article{ho2022entanglement,
  title={Entanglement-based quantum communication complexity beyond Bell nonlocality},
  author={Ho, Joseph and Moreno, George and Brito, Samura{\'\i} and Graffitti, Francesco and Morrison, Christopher L and Nery, Ranieri and Pickston, Alexander and Proietti, Massimiliano and Rabelo, Rafael and Fedrizzi, Alessandro and others},
  journal={npj Quantum Information},
  url = {https://www.nature.com/articles/s41534-022-00520-8},
  volume={8},
  number={1},
  pages={13},
  year={2022},
  publisher={Nature Publishing Group UK London}
}

@article{sen2003unified,
  title={Unified criterion for security of secret sharing in terms of violation of Bell inequalities},
  author={Sen, Aditi and Sen, Ujjwal and {\.Z}ukowski, Marek and others},
  journal={Physical Review A},
  url = {https://doi.org/10.1103/PhysRevA.68.032309},
  volume={68},
  number={3},
  pages={032309},
  year={2003},
  publisher={APS}
}

@article{moreno2020device,
  title={Device-independent secret sharing and a stronger form of Bell nonlocality},
  author={Moreno, MGM and Brito, Samura{\'\i} and Nery, Ranieri V and Chaves, Rafael},
  journal={Physical Review A},
  url = {https://doi.org/10.1103/PhysRevA.101.052339},
  volume={101},
  number={5},
  pages={052339},
  year={2020},
  publisher={APS}
}

@article{acin2012randomness,
  title={Randomness versus nonlocality and entanglement},
  author={Ac{\'\i}n, Antonio and Massar, Serge and Pironio, Stefano},
  journal={Physical review letters},
  url = {https://doi.org/10.1103/PhysRevLett.108.100402},
  volume={108},
  number={10},
  pages={100402},
  year={2012},
  publisher={APS}
}

@article{acin2007device,
  title={Device-independent security of quantum cryptography against collective attacks},
  author={Ac{\'\i}n, Antonio and Brunner, Nicolas and Gisin, Nicolas and Massar, Serge and Pironio, Stefano and Scarani, Valerio},
  journal={Physical Review Letters},
  url = {https://doi.org/10.1103/PhysRevLett.98.230501},
  volume={98},
  number={23},
  pages={230501},
  year={2007},
  publisher={APS}
}

@article{ekert1991quantum,
  title={Quantum cryptography based on Bell’s theorem},
  author={Ekert, Artur K},
  journal={Physical review letters},
  url = {https://doi.org/10.1103/PhysRevLett.67.661},
  volume={67},
  number={6},
  pages={661},
  year={1991},
  publisher={APS}
}

@article{nadlinger2022experimental,
  title={Experimental quantum key distribution certified by Bell's theorem},
  author={Nadlinger, David P and Drmota, Peter and Nichol, Bethan C and Araneda, Gabriel and Main, Dougal and Srinivas, Raghavendra and Lucas, David M and Ballance, Christopher J and Ivanov, Kirill and Tan, EY-Z and others},
  journal={Nature},
  url = {https://www.nature.com/articles/s41586-022-04941-5},
  volume={607},
  number={7920},
  pages={682--686},
  year={2022},
  publisher={Nature Publishing Group UK London}
}

@article{boreiri2023towards,
  title={Towards a minimal example of quantum nonlocality without inputs},
  author={Boreiri, Sadra and Girardin, Antoine and Ulu, Bora and Lipka-Bartosik, Patryk and Brunner, Nicolas and Sekatski, Pavel},
  journal={Physical Review A},
  url = {https://doi.org/10.1103/PhysRevA.107.062413},
  volume={107},
  number={6},
  pages={062413},
  year={2023},
  publisher={APS}
}

@article{bell1964einstein,
  title={On the einstein podolsky rosen paradox},
  author={Bell, John S},
  journal={Physics Physique Fizika},
  url = {https://doi.org/10.1103/PhysicsPhysiqueFizika.1.195},
  volume={1},
  number={3},
  pages={195},
  year={1964},
  publisher={APS}
}

@article{navascues2020genuine,
  title={Genuine network multipartite entanglement},
  author={Navascu{\'e}s, Miguel and Wolfe, Elie and Rosset, Denis and Pozas-Kerstjens, Alejandro},
  journal={Physical Review Letters},
  url = {https://journals.aps.org/prl/abstract/10.1103/PhysRevLett.125.240505},
  volume={125},
  number={24},
  pages={240505},
  year={2020},
  publisher={APS}
}

@article{pozas2023proofs,
  title={Proofs of network quantum nonlocality in continuous families of distributions},
  author={Pozas-Kerstjens, Alejandro and Gisin, Nicolas and Renou, Marc-Olivier},
  journal={Physical Review Letters},
  url = {https://doi.org/10.1103/PhysRevLett.130.090201},
  volume={130},
  number={9},
  pages={090201},
  year={2023},
  publisher={APS}
}

@misc{centeno2025significance,
      title={On the Significance of Intermediate Latents: Distinguishing Quantum Causal Scenarios with Indistinguishable Classical Analogs}, 
      author={Daniel Centeno and Elie Wolfe},
      year={2025},
      eprint={2412.10238},
      archivePrefix={arXiv},
      primaryClass={quant-ph},
      url={https://arxiv.org/abs/2412.10238}, 
}
\appendix

\section{Pearl interventions and Proof of Theorem \ref{thm:Pearl_from_partial} }\label{app: thm_1}
In this appendix we formalize the definition of Pearl interventions for quantum networks, and give the proof of Theorem  \ref{thm:Pearl_from_partial}. 
A Pearl intervention in a quantum scenario can be defined as:
\begin{defi}[Pearl intervention] \label{def: PearlInt} Given a network of $n$ parties,  $A_1,\dots,A_n$, with $m$ independent sources, $\Lambda_1,\dots,\Lambda_m$, where each source distributes states $\rho_i$. A Pearl intervention is defined relative to one of the $n$ parties of the network. Therefore, given a party $A_l\in V_O$, a Pearl intervention performed on $A_l$ is represented by the do-conditional $p(a_1,\dots,a_{l-1},a_{l+1},\dots,a_n|do(a_l))$,  given by 
\begin{equation}\label{eq: pearl_do}
P(a_1,\dots|do(a_l))=\text{Tr}\left(\bigotimes_{i}\rho_i\left[\left(\bigotimes_{ j\neq l}E_{a_j|Pa^O(a_j)}\right)\otimes \mathbf{1} \right]\right),
\end{equation}
where $p(a_1,\dots|do(a_l))$ takes statistics over all nodes $A_j\in V_O/\{A_l\}$. The definition essentially replaces the operator $E_{a_l|\text{Pa}^O(a_l)}$ by an identity $\mathbf{1}$, and leaves the remaining objects unchanged. 
%Equivalently, Eq.~(\ref{eq: pearl_do}) can be rewritten using the partial trace operation as 
%\begin{equation}
%   p(a_1,\dots|do(a_l))=\text{Tr}\left(\left[\text{Tr}_{A_l}\left(\bigotimes_i \rho_i\right)\right]\bigotimes_{j\neq l} E_{a_j|\text{Pa}^O(a_j)}\right).
%\end{equation}
\end{defi}
With this in mind, we now proceed to prove Theorem  \ref{thm:Pearl_from_partial}. It states: \emph{Consider a network  with sources $ \Lambda_1,\dots,\Lambda_m$, each distributing a system $\rho_i$, and parties $A_1,\dots,A_n$, performing measurements $E_{a_j|Pa^O(a_j)}$. Then any do-conditional on this network representing a Pearl-like intervention can always be inferred by a collection of latent splittings and the observational distribution.}

\emph{Proof.} First, let us define the sources  connected to $A_l$, $$\text{Pa}^L(A_l):=\{\Lambda_q\in V_L|\Lambda_q\rightarrow A_l\}.$$ Similarly the set of disconnected sources is $$\big(\text{Pa}^L(A_l)\big)^c:=\{\Lambda_q\in V_L|\Lambda_q\not\rightarrow A_l\}.$$ These two sets constitute a partition of all the original sources. As a consequence, we can write the global state as
       \begin{align}
        \rho_{obs}&=\bigotimes_{i\in V_L} \rho_i \nonumber\\ 
        &=\left(\bigotimes_{i\in \text{Pa}^L(A_l)}\rho_i\right)\otimes \left(\bigotimes_{i\notin \text{Pa}^L(A_l)}\rho_i\right).
    \end{align}
    Now consider the collection of latent splittings given by $\{(\Lambda_q,A_l)\}_{\Lambda_q}$ for $\Lambda_q
    \in \text{Pa}^L(A_l)$, which we further denote by  $\{(\Lambda_q,A_l)\}_{q}$ for $q\in \text{Pa}^L(A_l)$ for simplicity. 
     Here, each latent splitting will transform the state $\rho_q$ emitted by the source $\Lambda_q$ to the state $\hat{\rho}_{q,l}\otimes \text{Tr}_{A_l}(\rho_q)$, where $\hat{\rho}_{q,l}$ is a state prepared by $A_l$. The composition of all these interventions yields a global state $\rho_{int}$, given by 
    \begin{align}
        \rho_{int}&=\left(\bigotimes_{i\notin \text{Pa}^L(A_l)}\rho_i\right)\otimes \left(\bigotimes_{i\in \text{Pa}^L(A_l)}(\hat{\rho}_{i,l}\otimes \text{Tr}_{A_l}(\rho_i))\right).
    \end{align}
 Note that the states in  $\bigotimes_{i\notin \text{Pa}^L(A_l)}\rho_i$  are emitted by the sources in $\big(\text{Pa}^L(A_l)\big)^c$,  so the partial trace Tr$_{A_l}(\cdot)$ does not act over this product state,
   \begin{equation}
       \text{Tr}_{A_l}\left(\bigotimes_{i\in V_L} \rho_i \right) = \left(\bigotimes_{i\notin \text{Pa}^L(A_l)}\rho_i\right)\otimes \left(\bigotimes_{i\in \text{Pa}^L(A_l)}\text{Tr}_{A_l}(\rho_i))\right) .
   \end{equation}
   %Now, we may rewrite the state $\rho_{int}$ into a part that affects the system of the part $A_l$ and another part that does not affect $A_l$, 
   Therefore, we can rewrite $\rho_{int}$ as
 \begin{equation*}
      \rho_{int}= \text{Tr}_{A_l}\left(\bigotimes_{i\in V_L} \rho_i \right)  \otimes \left(\bigotimes_{i\in \text{Pa}^L(A_l)}\hat{\rho}_{i,l}\right). 
 \end{equation*}
 The share $\bigotimes_{i\in \text{Pa}^L(A_l)}\hat{\rho}_{i,l}$ is inside $A_l$'s laboratory, and thus, it can be subjected to any local operation performed by $A_l$. In particular, $A_l$ can apply a quantum channel $\mathcal{E}$ on its subsystems and prepare the final state  
 \begin{equation}
     \sigma_{A_l}:=\mathcal{E}\left(\bigotimes_{i\in \text{Pa}^L(A_l)}\hat{\rho}_{i,l}\right).
 \end{equation}
After all latent splittings and the application of the channel $\mathcal{E}$, the global state $\rho_{int}$ becomes
\begin{equation}
     \rho_{int}= \text{Tr}_{A_l}\left( \rho_{obs}\right)  \otimes \sigma_{A_l}.
\end{equation}
The statistics are given by 
\begin{align*}
    &P_{int}=\text{Tr}\left(\left[\text{Tr}_{A_l}\left(\rho_{obs}\right)  \otimes \sigma_{A_l}\right]\left(\bigotimes_{j\in V_O} E_{a_j |\text{Pa}^O(a_j)}\right)\right)\\
    &=\text{Tr}\left(\left[\text{Tr}_{A_l}\left(\rho_{obs} \right)  \bigotimes_{j\neq l} E_{a_j |\text{Pa}^O(a_j)}\right] \right)\text{Tr}\left( \sigma_{A_l} E_{a_l|\text{Pa}^O(a_l)}\right).
\end{align*}
In particular, $A_l$ by definition re-prepares the state $\sigma_{A_l}:= \text{Tr}_{V_O/\{A_l\}}(\rho_{obs})$, i.e. $A_l$ prepares $\rho_{obs}^{A_l}$ the partial state relative to $A_l$ of $\rho_{obs}$. In this case, by comparing with Def.~\ref{eq: pearl_do}, we can rewrite the $\rho_{int}$ as
\begin{equation}\label{eq: eq_thm_1}
    P_{int}(a_1,\dots,a_n)=P(a_1,\dots|do(a_l))P(a_l|do(\text{Pa}^O(a_l)).
\end{equation}

Note that Eq.~(\ref{eq: eq_thm_1}) is valid for an arbitrary part $A_l$ of the network. However, by definition, the network is represented by a DAG and thus, there must be at least one observable variable $A_{l^*}$, such that $\text{Pa}^{O}(A_{l^*})=\emptyset$. Note that, if such a node $A_{l^*}$ does not exist one can show that the causal structure must be cyclic. Therefore, we may use this general property of DAGs to estimate the do-conditional corresponding to $P(a_1,...|do(a_{l^*}))$, by using Eq. (\ref{eq: eq_thm_1}) as 
\begin{equation*}
    P_{int}(a_1,\dots,a_n)=P(a_1,\dots|do(a_{l^*}))P_{obs}(a_{l^*})
\end{equation*}
and, therefore, one can always use classical post-processing to $P_{int}, P_{obs}$ to obtain $P(a_1,\dots|do(a_{l^*}))$. 

Now, we can perform this procedure inductively. If we consider another node from $A_l\in V_O/\{A_{l^*}\}$ and $\text{Pa}^{O}(A_{l})=\emptyset$ we return to the previous case. However, if this is not the case, then there must be some node $A_r\in \text{Pa}^{O}(A_{l})$ such that $\text{Pa}^{O}(A_{r})=\emptyset$ and, thus, we can use Eq.~(\ref{eq: eq_thm_1}) to estimate the corresponding do-conditional of that intervention. Then, after estimating this do-conditional we are able to extend this to interventions on the \emph{children} variables of $A_r$, i.e. $\{X\in V_O|A_r\rightarrow X\}$. Using this procedure, we can cover all the observable variables in causal structure and, therefore, all possible Pearl-like interventions. \qed

\section{Equivalencies between latent splitting and Pearl interventions}
\label{app: partial_equal_Pearl}
Here, we will show two instances of Theorem \ref{thm:Pearl_from_partial} for the cases of the instrumental scenario and the unrelated cofounders (UC) scenario, shown in Fig. \ref{fig:app_B-DAG}. These two networks have already been explored in the context of Pearl interventions \cite{lauand2024quantum,poderini2024observational,gachechiladze2020quantifying} and here we show that collecting data from latent splittings is sufficient to obtain all data from do-conditionals.

First, let us consider the instrumental scenario, with observational distribution $P_{obs}(a,b|x)$, given by 
\begin{equation}
    P_{obs}(a,b|x)=\text{Tr}(\rho(E_{a|x}\otimes E_{b|a}))
\end{equation}

where $\rho$ is a quantum state shared between $A$ and $B$,
and the do-conditional statistics $P(b|do(a))$, are given by
\begin{equation}
    P(b|do(a))=\text{Tr}(\rho(\mathbf{1}\otimes E_{b|a}))
\end{equation}
Now, consider a latent splitting performed by either party on the source $\rho$, to obtain $\rho_{int}=\rho_A\otimes\rho_B$, where $\rho_A:=\text{Tr}_B(\rho)$ ans similarly for $\rho_B$, and the interventional statistics 
\begin{align}
    P_{int}(a,b|x)&:=\text{Tr}(\rho_A\otimes\rho_B(E_{a|x}\otimes E_{b|a}))\\
    &=\text{Tr}(\rho_AE_{a|x})\text{Tr}(\rho_BE_{b|a}).\nonumber
\end{align}
By using partial trace properties, we have the identities $\text{Tr}(\rho(E_{a|x}\otimes\mathbf{1}))=\text{Tr}(\text{Tr}_B(\rho)E_{a|x})$, and similarly $\text{Tr}(\rho(\mathbf{1}\otimes E_{b|a}))=\text{Tr}(\text{Tr}_A(\rho)E_{b|a})$. This allows us to derive 
\begin{equation}
     P_{int}(a,b|x)=P_{obs}(a|x)P(b|do(a)).
\end{equation}
Therefore, the interventional statistics reveal exactly the corresponding do conditionals $P(b|do(a))$. Conversely, one can always infer $P_{int}(a,b|x)$ from $P_{obs}(a,b|x)$ and $P(b|do(a))$.  

Now, we may consider the UC scenario, shown in Fig.~\ref{fig:app_B-DAG}, with $P_{obs}(a,b,c)$ given by
\begin{equation}
    P_{obs}(a,b,c)=\text{Tr}(\gamma \otimes \alpha(E_{a|b}\otimes E_{b}\otimes E_{c|b})),
\end{equation}

and the do-conditionals, given by 

\begin{equation*}
    P(a|do(b))=\text{Tr}(\gamma_A E_{a|b}),
\end{equation*}
and 
\begin{equation*}
    P(c|do(b))=\text{Tr}(\alpha_C E_{c|b}),
\end{equation*}
where $\gamma_A:=\text{Tr}_B(\gamma)$, and $\alpha_C:=\text{Tr}_B(\alpha)$.

Here, it is sufficient to consider the latent splitting performed by Alice on $\gamma$, denoted by $P_{int}^{\gamma}(a,b,c)$, and by Charlie on $\alpha$, denoted $P_{int}^{\alpha}(a,b,c)$, which are computed as
\begin{align*}
    P_{int}^{\gamma}(a,b,c):&=\text{Tr}(\gamma_{A}\otimes\gamma_B\otimes\alpha(E_{a|b}\otimes E_b \otimes E_{c|b}))\\
    &= \text{Tr}(\gamma_A E_{a|b})\text{Tr}(\gamma_B\otimes\alpha(E_b\otimes E_{c|b}))\\
    &=P(a|do(b))P_{obs}(b,c),
\end{align*}
and similarly $P_{int}^{\alpha}(a,b,c)=P_{obs}(a,b)P(c|do(b))$. Therefore, one can always infer the do-conditionals from latent splittings and vice-versa.
\begin{figure}[t]

\begin{minipage}[t]{0.48\columnwidth}
    \centering
    (a) Instrumental scenario
    \begin{tikzpicture}
        \draw (120 - 30: 1.cm) node[latent] (l1) {$\rho$};
        \draw (120 + 30: 1.4cm) node[var] (x) {$X$};
        \draw (240 - 30: .8cm) node[var] (a) {$A$};
        \draw (360 - 30: .8cm) node[var] (b) {$B$};

        \path[dir, forestGreen] (l1) edge (a) (l1) edge (b) (x) edge (a) (a) edge (b);
    \end{tikzpicture}

\end{minipage}
\begin{minipage}[t]{0.48\columnwidth}
    \centering
    (b) UC scenario
    \begin{tikzpicture}

        \draw (120 + 30: 1.6cm) node[latent] (l2) {$\gamma$};
        \draw (240 - 30: 1.6cm) node[var] (a) {$A$};
        \draw (240 + 30: .5cm) node[var] (c) {$B$};
        \draw (360 - 30: 1.6cm) node[var] (b) {$C$};
        \draw (360 + 30: 1.6cm) node[latent] (l3) {$\alpha$};

        \path[dir, forestGreen] (c) edge (a) (c) edge (b); 
        \path[dir, forestGreen] (l2) edge (c) (l2) edge (a);
        \path[dir, forestGreen] (l3) edge (b) (l3) edge (c);

    \end{tikzpicture}

\end{minipage}
\caption{\textbf{DAGs describing the instrumental and the UC scenarios.} }
\label{fig:app_B-DAG}
\end{figure}

\section{The interventional Fritz trick}
\label{app: Fritz_trick}
In this appendix, we provide a step-by-step derivation of  Eq.~(\ref{eq: Interventional_fritz}) from Eq.~(\ref{eq:chsh_2pint}) using the quantum strategy defined in the text:
$A$ and $B$ share the maximally entangled state $\ket{\psi^+}$, and they perform the optimal measurements for the bipartite Bell scenario %$A_0 =\sigma_z$, $A_1 =\sigma_x$ and $B_0 = (\sigma_x+\sigma_z)/\sqrt{2}$, $B_1 = (\sigma_z-\sigma_x)/\sqrt{2}$.  
such that the corresponding correlations are $ \braket{A_0B_0} = -1/\sqrt{2}$ and $ \braket{A_xB_y} = 1/\sqrt{2}$ for $xy\in \{01,10,11\}$.  Alice's inputs are determined by $\beta$, Bob's inputs are determined by $\alpha$ and Charlie outputs $c=\alpha\beta$.  The sources $\alpha$ and $\beta$ distribute classical bits, with $ p(\alpha = 1) = p(\beta = 1) = \epsilon$. Our goal is to calculate the expression
\begin{equation}
    S= (E_{\alpha\beta}+2)P_{obs}(c=1) - E^1_{obs}-E^1_{\alpha}-E^1_{\beta},
\end{equation}
where
\begin{align*}
     &E_{obs}^{c}=P_{obs}(a=b,c)-P_{obs}(a\neq b,c),\\
     &E_{\beta}^{c}=P^{\beta}_{int} (\hat a=b,c)-P^{\beta}_{int} (\hat a\neq b,c),\\
     &E_{\alpha}^{c}=P^{\alpha}_{int} (a=\hat b,c)-P^{\alpha}_{int} (a\neq \hat b,c),\\
     &E_{\alpha\beta }=P^{\alpha \beta}_{int} (\hat a=\hat b)-P^{\alpha \beta}_{int} (\hat a\neq \hat b),
 \end{align*}
 as defined in the main text.

First, let us begin by calculating $P_{obs}(c=1)$. Since $C$ outputs $c=\alpha\beta$ then  $c=1$ implies $\alpha=\beta=1$. Then 
\begin{equation}
    P_{obs}(c=1) = p(\alpha =1)p(\beta=1) =\epsilon^2.
\end{equation}
Now we can calculate $E^1_{obs}$. Again $c=1$, so $\alpha=\beta =1$. Notice that in the observational case, Alice's inputs are sent by the source $\beta$, while Bob's inputs are sent by the source $\alpha$. Therefore $x=y=1$, and we have that 

 \begin{align}
     E_{obs}^{1}& = P_{obs}(a=b,1)-P_{obs}(a\neq b,1)\nonumber\\
                & = \braket{A_1B_1}P_{obs}(c=1)\nonumber\\
                & = \frac{\epsilon^2}{\sqrt{2}}.
 \end{align}
Now consider the interventional terms. In the case of $E_{\alpha}^{1}$, the intervention is over $\alpha$. Then $B$ obtains his input from source $\hat \alpha$ with probability $p(\hat\alpha) = \epsilon$ (since $\hat \alpha$ and c are uncorrelated): 
\begin{align}
    E_{\alpha}^{1}& = P^{\alpha}_{int} (a=\hat b,1)-P^{\alpha}_{int} (a\neq \hat b,1)\nonumber\\
                  & = \sum_{\hat \alpha}p(\hat\alpha)\braket{A_1\hat{B}_{\hat\alpha}}P^{\alpha}_{int}(c=1)\nonumber\\
                  & = \big[\epsilon\braket{A_1\hat B_1} +(1-\epsilon)\braket{A_1\hat B_0}\big]\epsilon^2\nonumber\\
                   & = \frac{\epsilon^2}{\sqrt{2}}.
\end{align}
Analogously, when the intervention is performed on $\beta$, Alice obtains her input from $\hat\beta$. Therefore,
\begin{equation}
    E_{\beta}^{1} =  \frac{\epsilon^2}{\sqrt{2}} . 
\end{equation}
The last correlator $E_{\alpha\beta }$ corresponds to the scenario where the two sources $\alpha$ and $\beta$ have been intervened. In this case, both Alice and Bob receive their inputs from sources $\hat\beta$ and $\hat\alpha$ respectively. Therefore,

\begin{align}
    E_{\alpha\beta } &= p(\hat\alpha=1)p(\hat\beta=1)\braket{\hat A_1\hat B_1} \nonumber\\
                     &+ p(\hat\alpha=1)p(\hat\beta=0)\braket{\hat A_0\hat B_1}\nonumber\\
                     &+ p(\hat\alpha=0)p(\hat\beta=1)\braket{\hat A_1\hat B_0}\nonumber\\
                     &+ p(\hat\alpha=0)p(\hat\beta=0)\braket{\hat A_0\hat B_0}.
\end{align}
Substituting the probabilities $p(\hat\alpha=1)=p(\hat\beta=1)=\epsilon$,   $p(\hat\alpha=0)=p(\hat\beta=0)=1-\epsilon$ and the values of the correlators yields, 

\begin{equation}
    E_{\alpha\beta } =\frac{ \epsilon^2+2\epsilon(1-\epsilon)-(1-\epsilon)^2}{\sqrt{2}} 
    =\frac{-2\epsilon^2 +4\epsilon-1}{\sqrt{2}}.
\end{equation}
Finally, by substituting all  terms into the expression for  $S$ and rearranging, we recover Eq.~(\ref{eq: Interventional_fritz}),
\begin{equation}
    S  = \frac{4}{\sqrt{2}}( \epsilon-1)\epsilon^2-\frac{2}{\sqrt{2}} \epsilon^4 +2\epsilon^2.
\end{equation}

\end{document}